\documentclass[final,2p,times,twocolumn]{elsarticle}
\usepackage{amssymb}
\usepackage{amsmath}
\usepackage{amsfonts}
\usepackage{graphicx}
\usepackage{epsfig}
\usepackage{subfigure}
\usepackage{bm}
\usepackage{multirow}
\usepackage{array}

\newcolumntype{P}[1]{>{\centering\arraybackslash}p{#1}}

\journal{Computational Material Science}

\begin{document}

\begin{frontmatter}

\title{Director structures with dominant in-plane alignment in hybrid
planar films of biaxial nematic liquid crystals: A Monte Carlo study. }

\author[1]{B. Kamala Latha\corref{cor1}}

\author[1]{G. Sai Preeti\fnref{fn1}} 
\author[1]{K. P. N. Murthy\fnref{fn2}} 
\author[1,2]{ V. S. S. Sastry}
\address[1] {School of Physics,University of Hyderabad, Hyderabad 500046, India}
\address[2]{School of Engineering Sciences and Technology, University of Hyderabad, Hyderabad 500046, India}

\cortext[cor1]{corresponding author}
\fntext[fn1]{Present adress:Department of Physics, GITAM School of Technology,
GITAM Universiy, Hyderabad 500102, India}
\fntext[fn2]{Present adress: Manipal Centre for Natural Sciences, 
Manipal University, Manipal 576104, India}

\begin{abstract}
Equilibrium director structures in two thin hybrid planar films of 
biaxial nematics are investigated through Markov chain Monte Carlo 
simulations based on a lattice Hamiltonian model 
within the London dispersion approximation. While the substrates of the
 two films induce similar anchoring influences on the long axes of the 
liquid crystal molecules (viz. planar orientation at one end and 
perpendicular, or homeotropic, orientations at the other), they differ 
in their coupling with the minor axes of the molecules. In Type-A film 
the substrates do not interact with the minor axes at all (which is 
experimentally relatively more amenable), while in Type-B, 
the orientations of the molecular axes at the surface layer are 
influenced as well, by their biaxial coupling with the surface. Both 
films exhibit expected bending of the director associated with the ordering 
of the molecular long axes due to surface anchoring. Simulation results 
indicate that the Type-A film hosts stable director structures in the 
biaxial nematic phase of the LC medium, with the primary director lying 
in the plane of the film. High degree of this stable order thus  
developed could be of practical interest for potential applications. 
Type-B film, on the other hand, experiences competing interactions 
among the minor axes due to incompatible anchoring influences at the 
bounding substrates, apparently leading to frustration.    

\end{abstract}


\end{frontmatter}

\section{Introduction}
\label{}
The biaxial nematic phase ($N_{B}$) of liquid crystals (LC), predicted 
theoretically very early \cite{Freiser}, and realised experimentally 
in the past decade, in bent-core \cite{Acharya03, Acharya04, Madsen}, 
tetrapode \cite{Merkel, Figueirinhas05, Neupane}, and polymeric
\cite{Hessel, Severing} systems, is characterised by a primary director 
$\bm{n}$ and a secondary director $\bm{m}$ (perpendicular to $\bm{n}$).
Field-induced switching of the secondary director is envisaged to be 
faster than the primary director in the biaxial nematic phase, a fact 
which endows these nematics with a promising potential for use in fast 
switching electro-optic devices \cite{Lee07, Mamatha10}. 
The orthorhombic $N_{B}$ phases with $D_{2h}$ symmetry 
are suggested to be desirable \cite{Carsten} for ready applications. 

While the current experimental studies are still concerned with 
unambiguous confirmation of  macroscopic biaxiality \cite{Carsten, 
VanLe, Ostapenko}, theoretical studies have been more optimistic. 
 Modelling the Hamiltonian in terms of interactions among molecular
tensors, mean-field predictions within quadratic approximation envisage
systems which condense into liquid crystal phases with biaxial
symmetry \cite{Sonnet,matteis05A,Bisi06, Bisi07,matteis07}. These 
encompass  molecular structures with wide ranging symmetry \cite{Bisi08}. 
Computer simulations \cite{BerardiB} on the other 
hand have been playing a significant role  in investigating these models
systematically. Recent Monte Carlo studies based on this lattice model focus 
on the competing effects of different energy contributions in the Hamiltonian
on the ordering of the medium \cite{Dematteis08, Kamala14,Kamala15} in 
biaxial systems.  A molecular dynamics (MD) simulation of the bulk 
biaxial Gay-Berne fluid under the action of an electric field \cite
{Berardi08} has convincingly shown that the switching of the director,
associated with the minor molecular axes, is an order of magnitude faster 
than that of the director defined by the long molecular axes. 

In the biaxial nematic phase, a different pathway for fast switching 
between different  birefringent states (compared to conventional uniaxial 
LC systems) is possible because the birefringence can be changed by a 
rotation of the short axes which are thermally ordered, while the orientation 
of the long axes could be kept fixed \cite {Luck2}. A possible device 
configuration to achieve this objective is to use a film of biaxial liquid 
crystal confined in a planar cell with hybrid boundary conditions, wherein 
the geometry could constrain the orientation of the primary director (of the
long axes) in the biaxial nematic phase, leaving the secondary director 
( of one of the short axes) free for switching with an appropriate 
(in-plane) field. Studies on uniaxial hybrid films have established 
\cite{Pingsheng, Barbero3} that a bent-director configuration could be 
realised if the film thickness is greater than a critical thickness 
determined by the curvature elasticity of the medium and the surface 
interaction strength. Preliminary work on their biaxial counterpart was 
carried out earlier \cite{Gspthesis}.

     In this context, we investigated the equilibrium director structures 
 in two planar films of biaxial liquid crystals, in the uniaxial and 
biaxial phases of the medium. The anchoring influences of the two 
substrates comprising the cell are used to pin the orientation of the 
primary director [ordering direction of the major (long) axes of the 
molecules] near the two surfaces so as to result in a bent-director
 hybrid structure. The substrates can be chosen  either not to have 
 influence on the minor axes of the molecules (pure uniaxial coupling 
 with the substrate, say Type-A film), or to couple with the minor 
 axes as well (biaxial coupling with
the substrate, say Type-B film). We simulated director structures in both
these films based on Markov chain  Monte Carlo (MC) sampling technique,
constructing corresponding equilibrium ensembles. This paper reports our 
results examining the role of different anchoring conditions on the 
orientational ordering in the medium for potential applications. 
    
        In section II we introduce the lattice model of the medium, and
 details of the anchoring conditions of the two films. The MC simulation
are also discussed in this section. The equilibrium  director structures of
the film in the two nematic phases obtained from the computations are  
 depicted and discussed in section III. We also examine the effect of 
 varying the cell thickness, as well as of the relative anchoring strengths 
 at the two substrates, on the director structures in these films. The last
  section summarizes our conclusions. 
  
 \section{Model and Simulation details}
\label{•}
\begin{figure}
\centering
\includegraphics[scale=0.2]{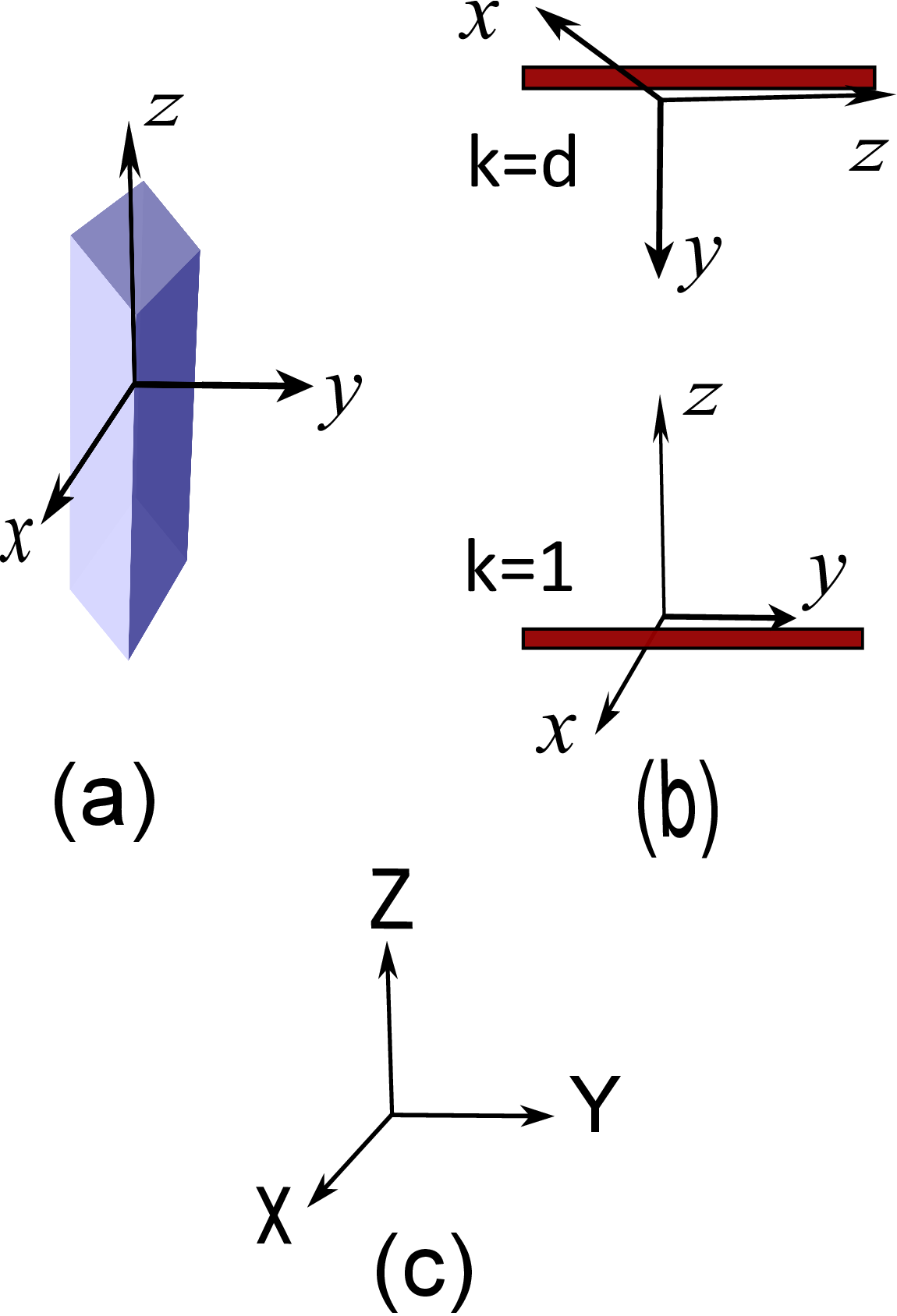}
\caption{Schematic illustrations of (a) Typical biaxial molecule (b)
anchoring directions at the two substrates of the film: homeotropic
at the lower (k=1) substrate and planar at the top (k=d)
(c) orientation of the laboratory  frame relative to the film.}
\label{fig:1}
\end{figure}

We consider a planar hybrid film comprising of LC molecules with $D_{2h}$
symmetry. We assign the right-handed triad $ \lbrace X,Y,Z\rbrace$ to 
represent the laboratory-fixed frame and $\lbrace \textit{x, y, z}\rbrace$
 to represent the molecular-fixed frame.  We let $\textit{z}$ direction 
represent the molecular long axis, while the other two (minor) axes are 
represented by $\textit{y}$ and $\textit{x}$. The film is obtained by 
confining the biaxial molecules between two planar substrates taken
to be in the X-Y plane. Fig.~\ref{fig:1} shows the schematics of a 
biaxial molecule, orienting influences at the two substrates of the 
planar film, and the reference axes of the laboratory.   
The orientational interactions between LC 
molecules, relevant to the present study, are conveniently accounted for, 
by adopting a lattice Hamiltonian model wherein the molecules located 
at the lattice sites are represented by unit vectors in the
$ \lbrace \textit{x},\textit{y},\textit{z} \rbrace$ frame specifying the 
individual molecular orientations. Within this lattice description, in
 a film of thickness $d$ the substrate planes are positioned at 
$Z=0$ and $Z=d+1$ (lattice units). The influence of the 
substrates is simulated by introducing two bounding layers of molecules
contained in these planes with the designated, but fixed, orientations, 
referred to in the literature as ghost molecules \cite{Pasini1}. 
 The anchoring conditions in both the films are such that the 
long axes of the molecules are hybrid-aligned (planar orientation parallel
to say, $Y$-axis at one substrate and homeotropic at the other, parallel to 
$Z$-axis). For adequately thick films, the primary director $\bm{n}$ is 
bent satisfying the two incompatible boundary conditions. We distinguish
two scenarios: (a) in Type-A  film, the ghost molecules interact with 
the LC molecules in the surface layer anchoring only their long axes, thus 
implying that the substrate hosts only cylindrically symmetric rod-like LC
constituents, and (b) in Type-B film the ghost molecules themselves have
$D_{2h}$ symmetry and interact through a Hamiltonian model appropriate 
to the biaxial system. These correspond to two qualitatively distinct 
chemical treatments of the anchoring substrates.

         In Type-A film, the molecular $\textit{z}$-axes are anchored 
along the Z-direction in the surface layer near, say, the lower substrate 
of the cell (k=1), while they are kept planar (say, parallel to Y-axis) 
in other surface layer, near the substrate (k=$\textit{d}$). As the substrates 
do not interact with the minor axes of the molecules in this film, 
they are not, $\textit{a priori}$, oriented
in any specific direction, until guided to equilibrium conditions
by the intermolecular and substrate interactions.  

             In the Type-B film, we impose anchoring conditions of equal 
strength on all the three molecular axes at each of the two substrates, and
use the biaxial Hamiltonian model to account for their interactions with 
the substrates. The boundary conditions on the film at the two surfaces are
summarized as (see Fig.~\ref{fig:1}):\\
\begin{align*}                    
\text{Substrate 1 at Z=0} &: z \parallel Z, y \parallel Y, x \parallel X \\
\text{Substrate 2 at Z=d+1} &:  z \parallel Y, y \parallel Z, x \parallel X
\end{align*}

\subsection{Model Hamiltonian}
The biaxial LC molecules are assumed to interact through a pair-wise 
additive lattice Hamiltonian  within the London dispersion 
approximation \cite{Pdz95}, expressed in terms of generalised Wigner rotation 
matrices as: 
\begin{equation}
\begin{split}
U(\omega_{ij}) =-\epsilon_{ij} \lbrace P_{2}(\cos (\beta_{ij}))+2 \ \lambda_{d}\ [R^{2}_{02}(\omega_{ij})+ R^{2}_{20}(\omega_{ij})]\\
			    +4 \  \lambda_{d}^{2} \  R^{2}_{22}(\omega_{ij})\rbrace
\label{eqn:1}
\end{split}
\end{equation}
where $\epsilon_{ij}=\epsilon$ sets the energy scale, and is  used to define 
the reduced temperature, $\omega(\alpha,\beta,\gamma)$ is the set of 
Euler angles which specify the rotations to be performed in order to 
bring the reference frame of two molecules $\textit{i}$ and $\textit{j}$ in 
coincidence, $R^{2}_{mn}$ are symmetrized Wigner functions, 
 $P_{2}(\cos(\beta_{ij}))$ is the second Legendre polynomial and 
$\lambda_{d}$ quantifies the biaxial interaction between the molecules. The 
average values of $R^{2}_{mn}$ define the order parameters of the medium 
in the nematic phases. These are: the uniaxial order $<R^{2}_{00}>$ 
(along the primary director), the phase biaxiality $<R^{2}_{20}>$, and 
the molecular contribution to the biaxiality of the medium  $<R^{2}_{22}>$, 
and the contribution to uniaxial order from the molecular minor axes 
$<R^{2}_{02}>$ \cite{Pdz95}.  
 For simulation purposes, the above Hamiltonian is recast in the 
 Cartesian form , as
 
 \begin{equation}
U=-\epsilon\lbrace\dfrac{3}{2}V_{33}- \sqrt{6} \ \lambda_{d} \ (V_{11}-V_{22})+\lambda_{d}^{2}\ (V_{11}+V_{22}-V_{12}-V_{21})-\dfrac{1}{2}\rbrace .
 \label{eqn:2} 
 \end{equation}
 
Here, $V_{ab}= (u_{a}.v_{b} )^2$,  and the unit vectors $u_{a},v_{b}$, 
$[a, b = 1, 2, 3]$, are the three axes of the  two interacting neighbouring 
molecules. $\lambda_{d}$ sets the relative importance of the biaxial 
interaction in the Hamiltonian, while $\epsilon$ (set to unity in the
 simulations) defines the temperature scale 
 $(T^{'}= \dfrac{k_{B} T^{*}}{\epsilon})$ , where $T^{*}$ is the 
 laboratory temperature in Kelvin.

This model in a bulk system was studied extensively both through 
mean-field analysis and MC simulations based on Boltzmann sampling
methods. The value of $\lambda_{d}$ for the present study is kept at 
0.35, keeping in view the high degree of biaxiality it induces, as well
as the convenience of a wider biaxial nematic range of temperature made 
available for our study \cite{Pdz95}.

\subsection{Simulation Details}
A planar hybrid film of (lattice) dimensions $15\times15\times{d}$ 
($\textit{d}$ = 6, 8, 10, 12 layers) is considered in the present 
work. Periodic boundary conditions are applied along the laboratory X and
 Y directions, so as to minimize finite size effects. The anchoring 
conditions applied at the two substrates (contained in the X-Y plane)
depend on the specific choice of the film (Type A or B), and their 
relative strengths are chosen as desired. We index the LC layers with
 $\textit{k}$, starting from
the substrate imposing homeotropic anchoring influence on the long molecular
axes. The LC ghost molecules in the substrate layers (which are adjacent to 
the two bounding layers of the LC medium) do not participate in the 
Monte Carlo dynamics. The interaction strengths of the long molecular 
axes at the two substrates are represented by $\epsilon_{1}$ 
and $\epsilon_{d}$. $\epsilon_{1}$ is set equal to one and 
$\epsilon_{d}$  is varied relative to $\epsilon_{1}$, taking values 
0.1, 0.2, .....1.0. The strong anchoring case corresponds to 
$\epsilon_{1} = \epsilon_{d}= 1$ at the substrates. All simulations except 
those involving the explicit variation of the relative anchoring strengths (Section
3.4), are carried out under strong anchoring conditions.  The temperature $T^{'}$ 
(in dimensionless units) is set by the coupling strength $\epsilon$ 
in the Hamiltonian (Eqn.\ref{eqn:1}). 

The simulation always starts from an initial (random) configuration and
the Markov chain dynamics is effected by random moves in the configuration space, 
accepted or rejected as per the Metropolis algorithm. This Monte Carlo
procedure ensures that the system is attracted eventually to its basin of
equilibrium states consistent with the simulation conditions. Attainment
of equilibration is borne out by the stationarity of the properties of the 
system, like energy; in equilibrium their fluctuations are centered about
a constant mean value. In each simulation, the reduced temperature is 
varied from 2 to 0.05 in steps of 0.005, and at each temperature the 
film has been found to be well equilibrated after $5\times 10^{5}$ 
lattice sweeps (attempted moves over all the sites). Data 
from the resulting canonical ensembles 
  are collected over a production run of $5\times 10^{5}$ lattice sweeps 
 yielding acceptably small sampling  errors.
The physical properties computed are the average values of energy $E$, the 
specific heat $C_{v}$, the order parameters $R^{2}_{00}$, $R^{2}_{02}$, 
$R^{2}_{20}$, $R^{2}_{22}$ \cite{Pdz95,Low} and their susceptibilities. 
(The susceptibility of the order parameter, say X, at temperature $T^{'}$  is 
computed as $(< X^2 > - < X >^2)/T^{'} $). The   
 layer-wise orientation of the local directors corresponding to the 
 molecular $\textit{z}$-axes (orientation averaged 
over the layer) with respect to the laboratory Z-axis (polar angle $\theta$, 
layerwise) as well as the angle made by this local layer-wise director with
 respect  to laboratory X-axis (azimuthal angle $\phi$, layerwise) are also 
 presented. We computed both layer-wise properties (to examine this director structure 
 and its relative changes) as well as the bulk film properties (averaged over
the sample), as a function of temperature, for a fixed  relative 
anchoring strength and layer thickness. The simulations are then repeated 
by varying the thickness effecting the length scale of the system, and 
also varying the anchoring strength ($\epsilon_{d}$) to look for possible 
anchoring-induced transitions among the different director structures (Sections
3.3 and 3.4).

Errors from finite sizes of the MC samples are estimated employing resampling 
methods based on the jackknife (JK) algorithm \cite{Quenonille,Shao}. For 
this purpose, the total number of sampled data ( $5\times 10^{5}$ ) is 
divided into 1000 subsets, each consisting of 500 contiguous microstates. 
The physical quantities of interest are averaged over each of these 
subsets, yielding $10^{3}$ data points for each variable. The JK algorithm 
is applied  to this reduced data set, to compute the averages and sampling 
errors. This resampling technique is known to reduce artefacts that could
 arise due to probable correlations in the original sampled data.  

\section{Results and Discussion} 
\label{.}
\begin{figure}
\centering
\includegraphics[width=0.38\textwidth]{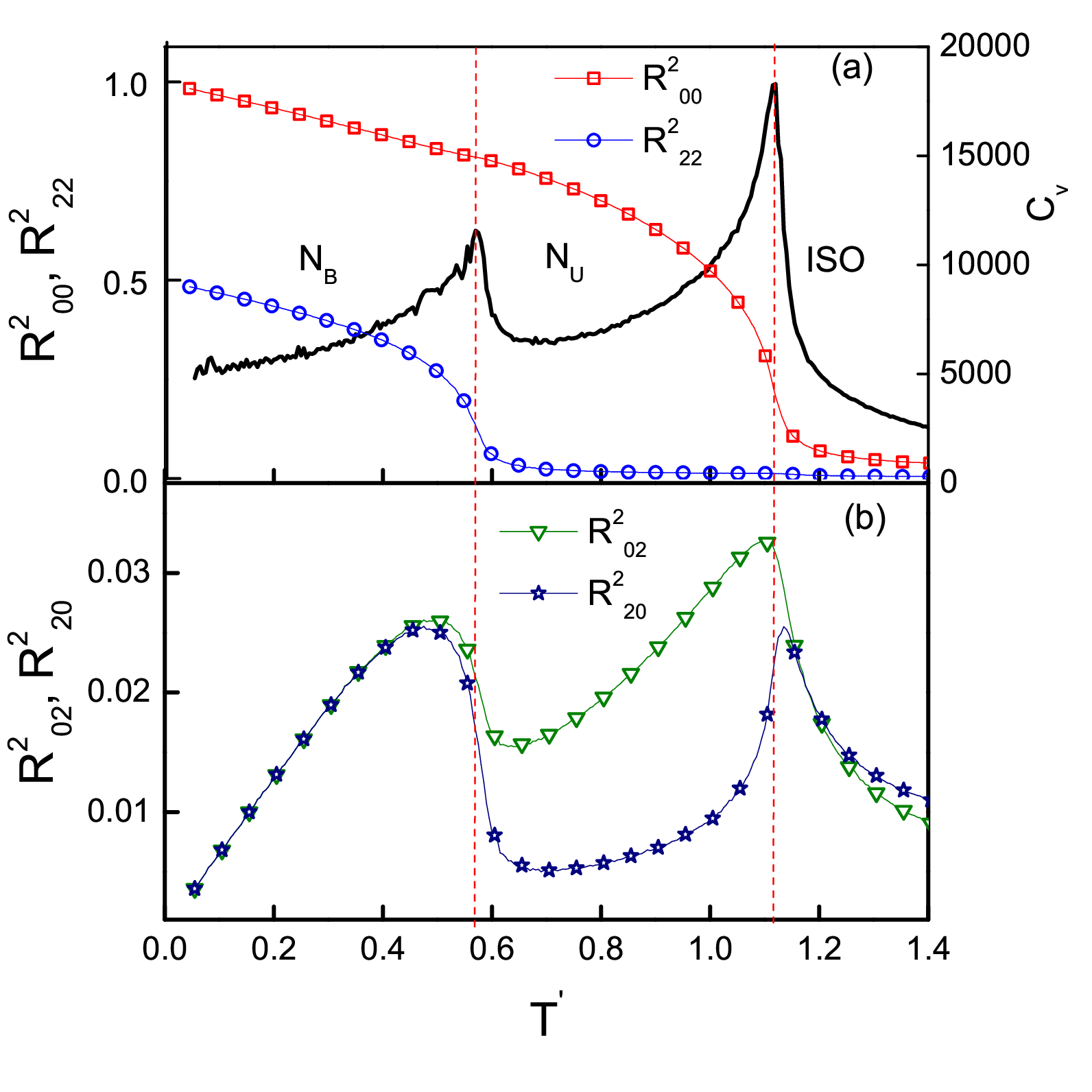}
\caption{Variation with temperature of system properties a bulk biaxial 
film  ($15\times15\times15$): specific heat $C_{v}$ peaks mark the phase 
 transitions $I-N_{U}$ and $N_{U} - N_{B}$. The temperature variation of the  
order parameters(a) $R^{2}_{00}$, $R^{2}_{22}$ (b) $R^{2}_{02}$ and $R^{2}_{20}$
 can be clearly observed.
 }
\label{fig:2}
\end{figure}

Fig.~\ref{fig:2}(a) depicts the variation of specific heat 
and the  two major order parameters in a bulk  biaxial fluid
(without confinement) of comparable dimensions ($15\times15\times15$) computed 
based on MC simulations under periodic boundary conditions, using the 
above dispersion Hamiltonian model at $\lambda_{d}$=0.35. 
The two peaks in the specific heat at $T_{1}$ = 1.123 and 
$T_{2}$ = 0.571 indicate the high temperature $I - N_{U}$ and 
low temperature $N_{U} - N_{B}$ phase transitions \cite{Pdz95}. 
The phases, as identified by the Metropolis-based Boltzmann sampling
procedure employed here,  are marked by the progressive growth of the uniaxial 
order $R^{2}_{00}$ and the biaxial order $R^{2}_{22}$, as a function of 
temperature, bringing out the onset of the uniaxial phase $N_{U}$ and low 
temperature biaxial phase $N_{B}$. The variation of $R^{2}_{02}$ and $R^{2}_{20}$
with temperature, as shown in Fig.~\ref{fig:2}(b), is the expected variation in 
the bulk system.  Figs.~\ref{fig:3} - \ref{fig:8} show
the simulation results for Type-A and Type-B films of thickness $d=8$ under 
strong anchoring conditions ($\epsilon_{1} = \epsilon_{d}= 1$). 
Fig.~\ref{fig:3} shows the onset of the
ordering at the $I - N_{U}$ and $N_{U} - N_{B}$ transitions (system 
properties) in the Type-A film. Fig.~\ref{fig:4}  provides variation of all the 
 four order parameters and their susceptibilities in the  Type-A film, 
 averaged over the sample. The qualitative difference between the thermal 
 evolution of these order parameters in the bulk and Type-A film (in 
 particular $R^{2}_{00}$) may be noted for further discussion later. The 
 director structure in the  film is better appreciated 
by focussing on the ordering tensor of the molecular $\textit{z}$-axes and 
examining its degree of ordering in each layer  and the orientation
 of the corresponding local layer director with respect to laboratory axes, specified 
 by their $\theta$ and $\phi$. These layer-wise properties
 are shown for Film-A in Fig.~\ref{fig:5}. Similar data were obtained for 
 Film-B as well, and are represented in Figs.~\ref{fig:6} - \ref{fig:8},
 respectively. It is to be noted that layer-wise angles of the local directors 
 are obviously not meaningful in the isotropic phase, and hence such 
 information in  these figures (Figs.~\ref{fig:5} and \ref{fig:8}) 
 (generated automatically  during computations) are to be ignored; such 
 data are relevant only in  the ordered phases. 
 
\begin{table*}
\caption{Typical errors of computation in the two films, as estimated from a 
single MC production run of $10^5$ data points, with the
jack-knife method}
\begin{center}
\begin{tabular}{| c | c | c | c | c | }
 \hline
 \centering Physical variable &\multicolumn{4}{|c|}{Average values with error estimates} \\
 \cline{1-5}
 &\multicolumn{2}{|p{3cm}|}{\centering $N_{U}^{'}$ phase ( T = 0.7 ) } &  \multicolumn{2}{|p{3cm}|}{\centering $N_{B}$ phase ( T = 0.2 ) }\\
\cline{2-5}
\centering & Type-A  & Type-B & Type-A &Type-B\\
\cline{1-5}
\centering Energy per lattice site & $-2.398 \pm 0.0003$& $-2.415 \pm 0.0002$ & $-3.484 \pm 0.0001$ & $-3.492\pm 0.0001$\\
\hline
\centering $R^{2} _{00}$  & $0.567 \pm 0.0003$ & $ 0.535 \pm 0.0002$ & $0.903 \pm 0.0001$ & $0.837 \pm 0.0004$\\
\hline
\centering $R^{2} _{22}$  & $0.090 \pm 0.0006$ & $0.042 \pm 0.0003$ & $0.387 \pm 0.0001$  & $0.434 \pm 0.0001$ \\
 \hline
\end{tabular}

\end{center}  
\label{tab:table1}
\end{table*}

 Error estimates of different physical properties (energy, $R^{2}_{00}$ 
 and $R^{2}_{22}$) are presented in Table 1 for both the films. Two representative
 temperatures as indicated are chosen in the two nematic phase, and the averages
 as well as errors (from the JK algorithm) are shown in the Table. These 
 are sampling errors from a singe production run, representing standard
 deviation of the corresponding MC mean values. As may be seen, they are too
 small to be shown along with symbols (representing the data points) in the 
 figures. These values indicate that the corresponding MC estimates are 
 robust in terms  of their reliability. We further tested whether different 
 trajectories originating from distinct random initial configurations (say,
 ten such initial conditions) would lead to the comparable values of 
MC estimates within these errors, to confirm the uniqueness (or otherwise)
of the free energy minimum (with respect to the different observables).
We carried out such MC simulations on both these films over the temperature
region and estimated the standard deviation of the MC values over these 
trajectories. These show that in Type-A film the error arising from 
scatter of the MC averages from different simulations are comparable, and 
are of same order of representative JK errors from a single MC simulation. This 
indicates a unique, fairly deep, free energy minimum in Type-A in both the
 nematic phases. Such a  comparison in Type-B film shows that the free energy 
 minimum seems to be  similar in the intermediate phase, but not in the biaxial phase.
 The larger scatter from different trajectories in the biaxial phase betrays a shallow
 free energy minimum with respect to the order parameter; there appear to be several  local minima, forming basins of attraction for different starting initial
 conditions. These observations on errors are relevant for later discussion.

\subsection{Type-A Film}

\begin{figure}
\centering
\includegraphics[width=0.38\textwidth]{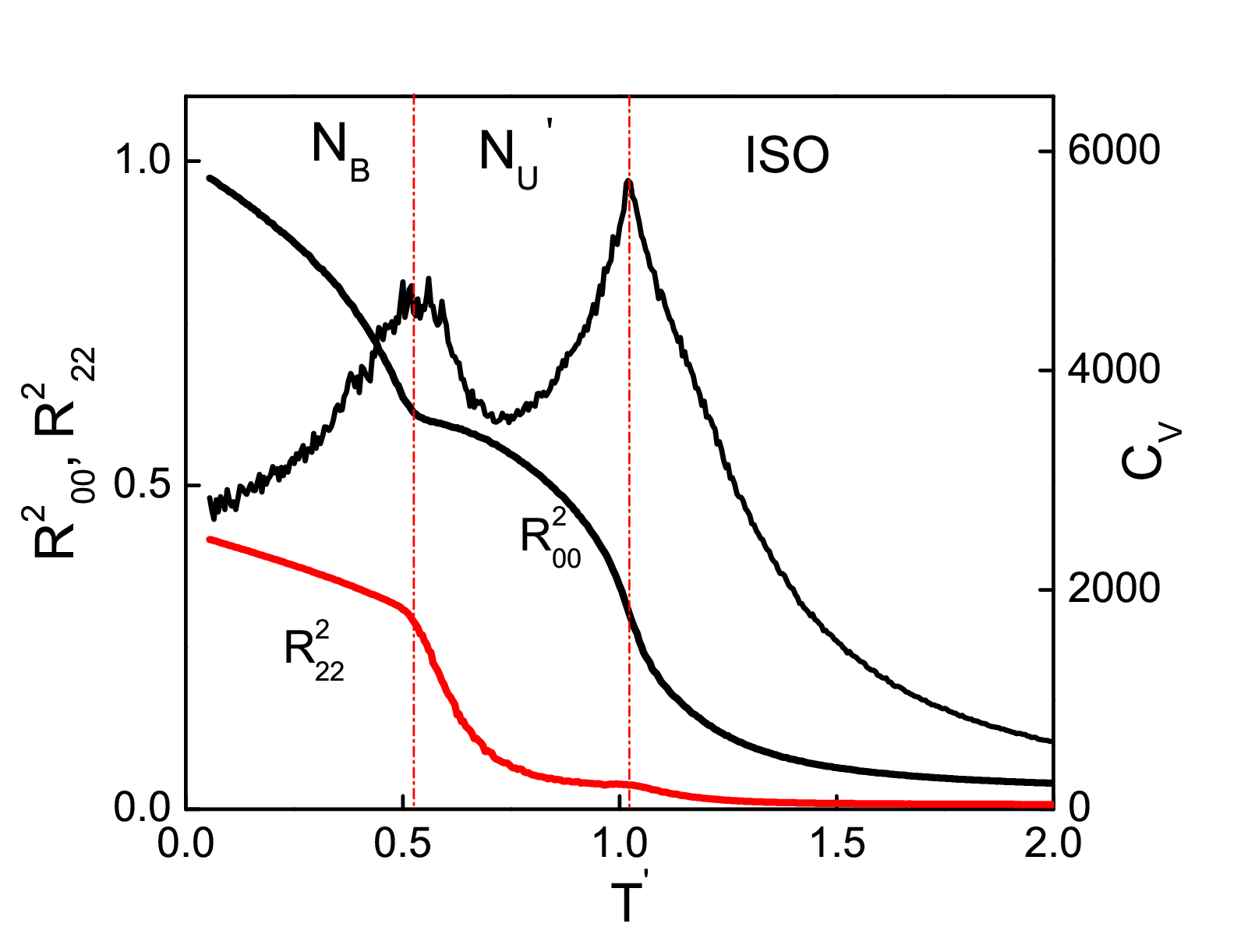}
\caption{Variation with temperature of system properties in Type-A film. 
Specific heat $C_{v}$ peaks mark the phase transitions $I-N_{U}^{'}$ and 
$N_{U}^{'} - N_{B}$. The  bent - director structure of uniaxial order parameter 
$R^{2}_{00}$ in the uniaxial phase is visible clearly.}
\label{fig:3}
\end{figure}

We note at the outset that in this film the substrate interacts with only
the long axes of the molecules, while LC molecules themselves have biaxial 
interaction among them. This system (Fig.~\ref{fig:3}) undergoes two transitions at 
$T_{1}=1.021$, and the second at $T_{2}=0.525$, both being lower than 
the corresponding temperatures in the bulk system (see Fig.~\ref{fig:2}). 
Keeping in view the significant
biaxial order developed in the films even in the intermediate phase and 
recognising that its origin is the spatial (layer-wise) inhomogeneities 
in the ordering tensors of individual layer structures ( with respect to 
the dominant director from the ordering of the molecular $\textit{z}$-axes), we 
refer to this as $N_{U}^{'}$ phase (rather than $N_{U}$), in 
line with the nomenclature in vogue to denote such uniaxial phases 
hosting inhomogeneous regions of biaxial order \cite{Peroukidis}.

\begin{figure}
\centering{
\subfigure[]{\includegraphics[width=0.38\textwidth]
{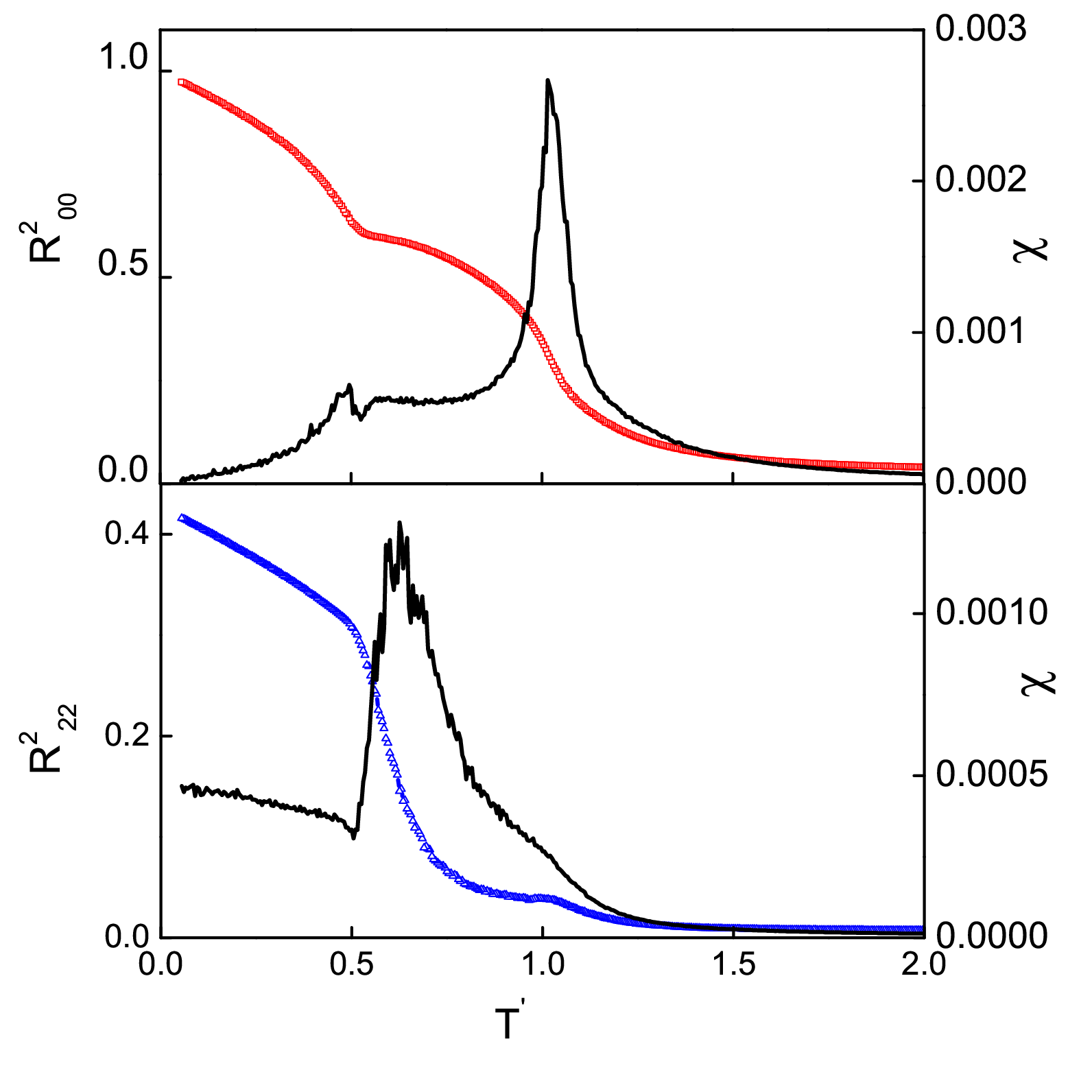}
\label{fig:4a}}
\subfigure[]{\includegraphics[width=0.38\textwidth]
{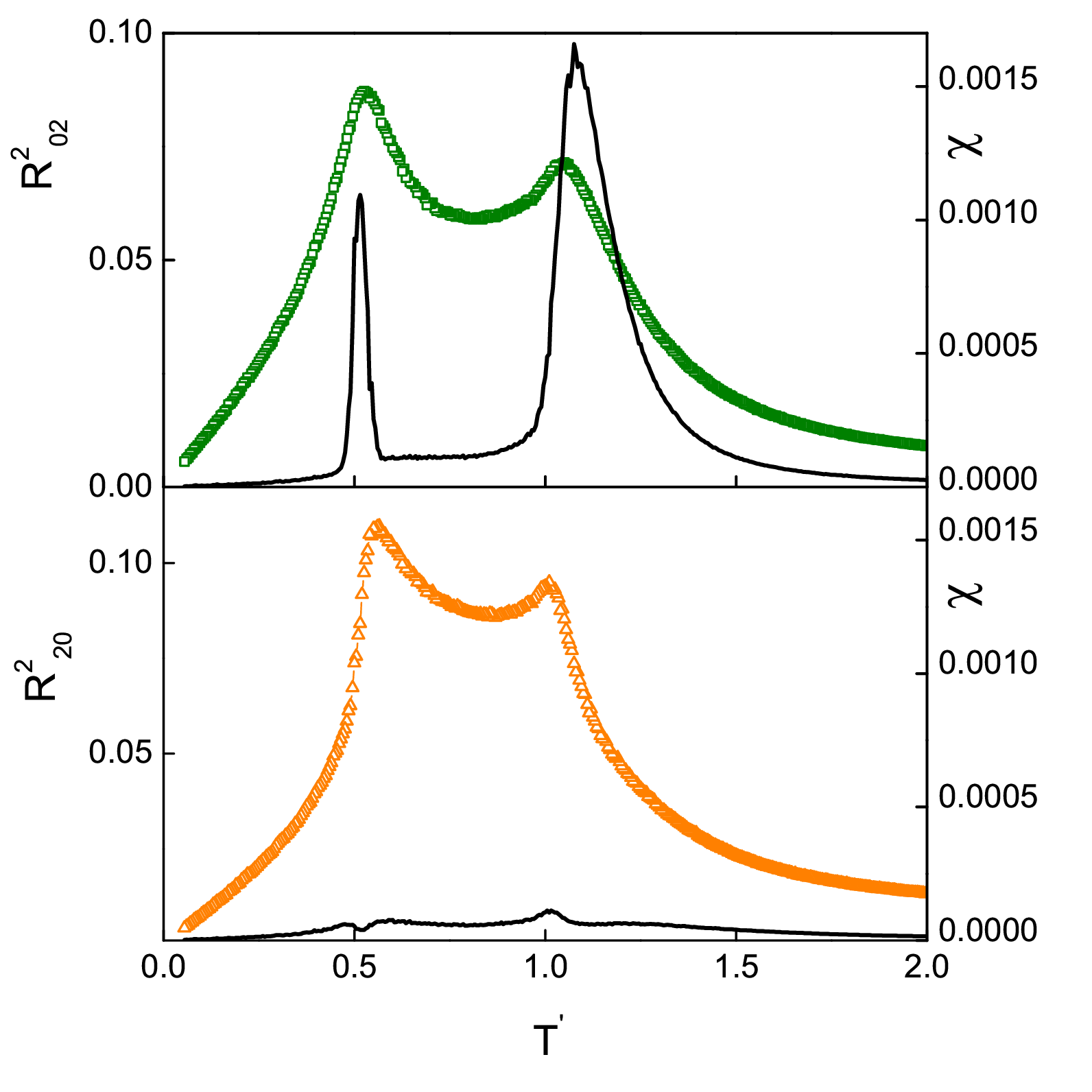}
\label{fig:4b}}
}
\caption{Variation with temperature of the four system order parameters 
(lines with symbols) and their susceptibilities (continuous lines) in 
Type-A film (a) $R^{2}_{00}$ and  $R^{2}_{22}$ (b) $R^{2}_{02}$ and 
$R^{2}_{20}$. }
\label{fig:4}
\end{figure}

 Figs.~\ref{fig:4a} and \ref{fig:4b} show the four order parameters of 
 the system along with their susceptibilities.  It is observed from 
 Fig.~\ref{fig:4a} that the $R^{2}_{00}$ susceptibility profile, which 
 is a measure of the fluctuations in the long-range order 
 of the primary director, shows two peaks. The larger peak (at $T_{1}$) 
 corresponds to rapid increase of $R^{2}_{00}$  in the uniaxial nematic 
 phase. The $R^{2}_{00}$ curve displays a bent-director
 structure and attains a maximum value of 0.6 in this phase. The smaller 
 peak close to $T_{2}$ signals a sudden change in the slope of the 
$R^{2}_{00}$ curve leading to a steady increase of the dominant (uniaxial) 
order towards a maximum value of unity, deep in the biaxial phase. The 
susceptibility of the biaxial order, on the other hand, shows a 
 single broad peak at a temperature slightly higher than $T_{2}$, 
 signalling the onset of biaxial order in this phase.

       Contribution to the  ordering along the uniaxial director originating 
from molecular  minor axes, $ R^{2}_{02}$ and the phase biaxiality parameter 
arising  from the non-cylindrical distribution of the molecular long axes 
around the primary director, $R^{2}_{20}$ are shown in Fig.~\ref{fig:4b}. As 
 compared to their values in a  bulk system (Fig.\ref{fig:2}),  these parameters have larger values in the  $N_{U}^{'}$ phase, - a manifestation of the geometric confinement. 
          
\begin{figure}
\centering{
\subfigure[]{\includegraphics[width=0.38\textwidth]{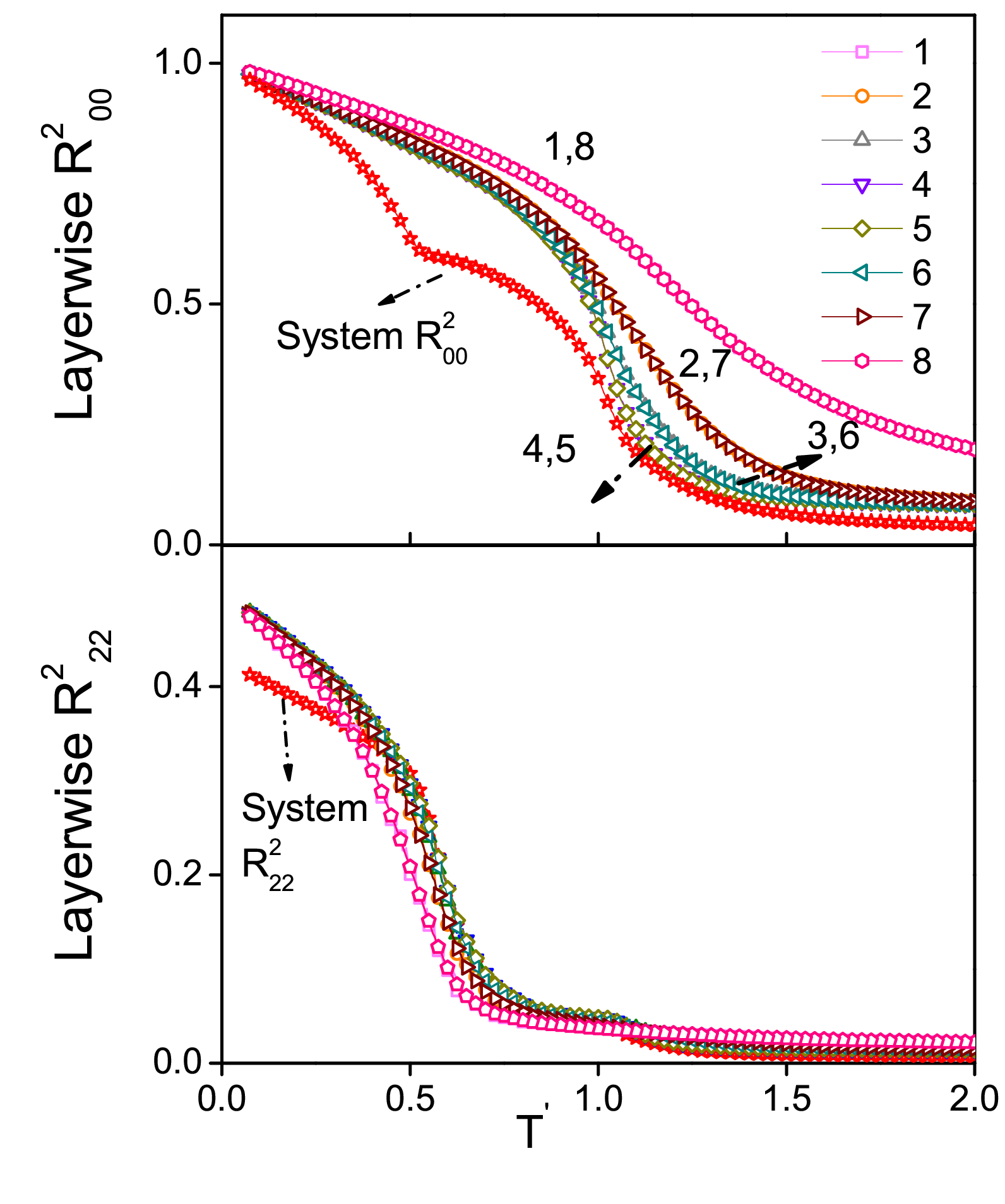}
\label{fig:5a}}
\subfigure[]{\includegraphics[width=0.38\textwidth]{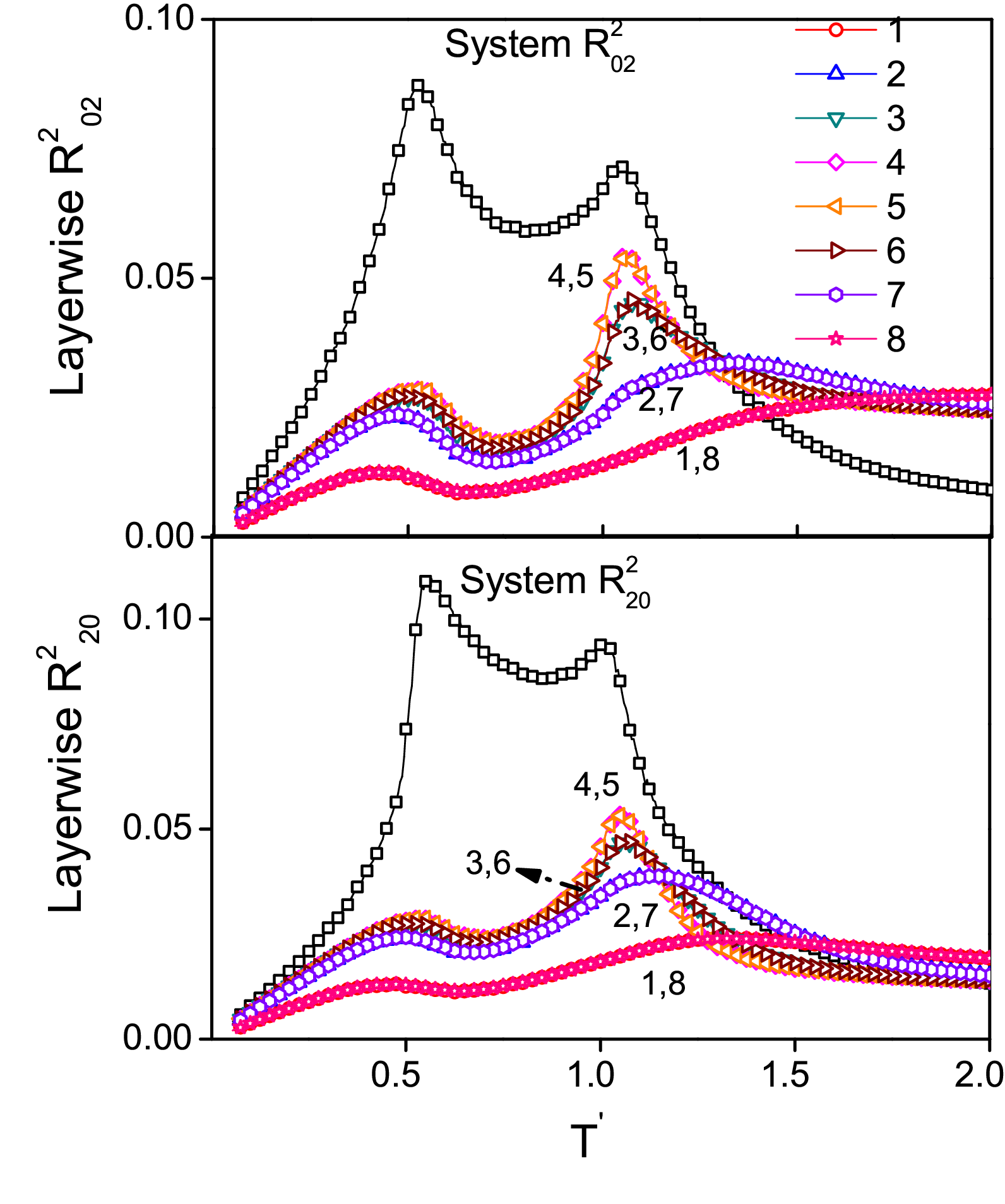}
\label{fig:5b}}
\subfigure[]{\includegraphics[width=0.35\textwidth]{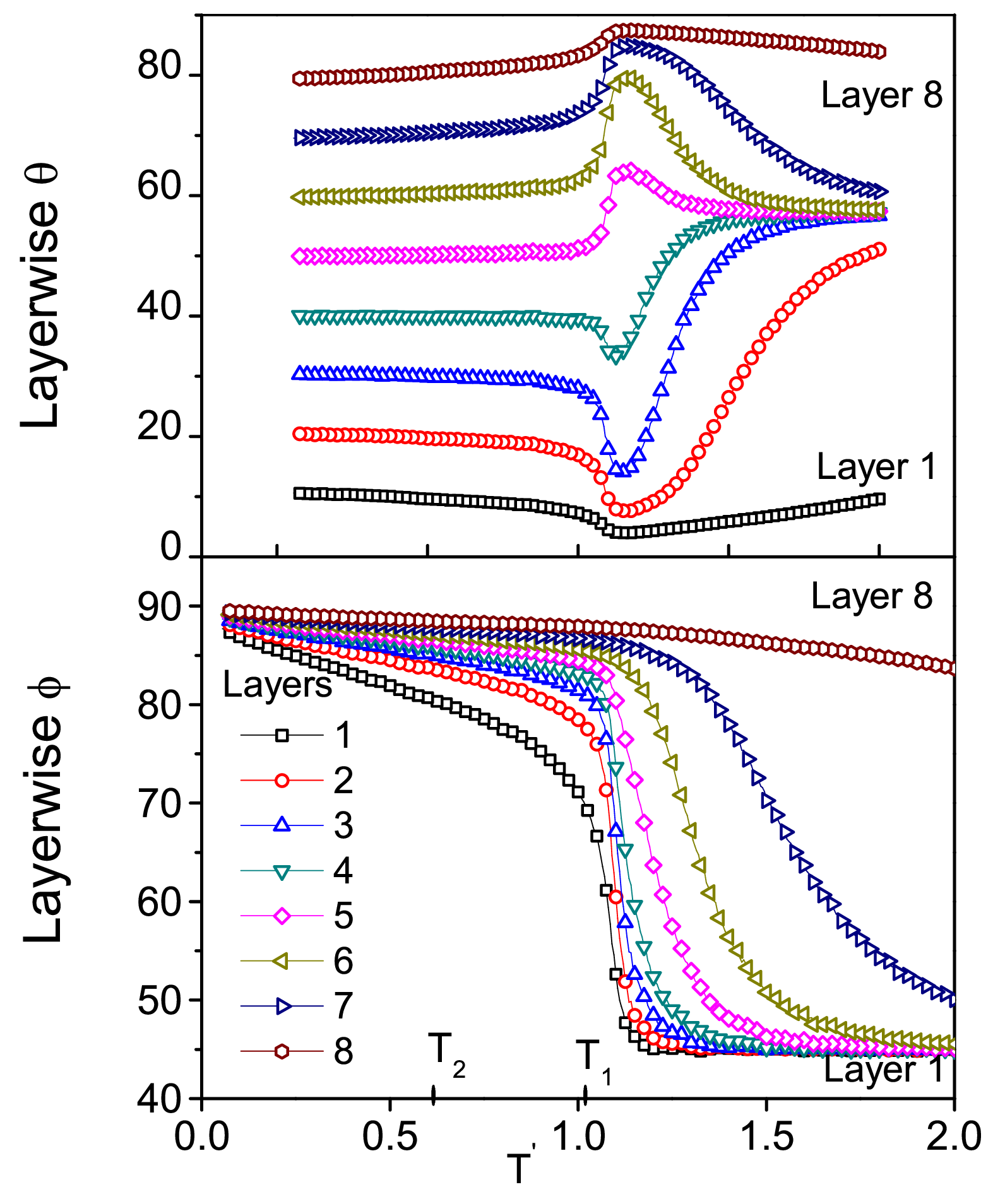}
\label{fig:5c}}
}
\caption{Layer-wise variation of different properties of the 
Type-A film, with temperature: (a) $R^{2}_{00}$ and $R^{2}_{22}$ (b) 
$R^{2}_{02}$ and $R^{2}_{20}$ (c) Angles $\theta$ and $\phi$ (see text). }
\label{fig:5}
\end{figure}

The layer-wise variations (for the layer index $\textit{k} = 1$ to $d$) 
of properties connected with director structures, plotted along with 
the corresponding bulk values of the film (sample averages) are shown 
in Figs.~\ref{fig:5a} - \ref{fig:5c}. It is
observed from Fig.~\ref{fig:5a} that layer-wise $R^{2}_{00}$ values vary 
smoothly, unlike the abrupt jump at $T_{2}$ exhibited by the 
sample average.  Further, the order values in all the layers
asymptotically reach the maximum value of 1.0 at the lowest temperature. 
The variation also shows that the middle layers 
($\textit{k}$ = 4 and 5), being the least influenced by the substrate 
boundaries, are most effective in contributing to the critical onset of the 
order at the transition. The sample average and layer-wise 
behaviour of biaxial order shows that in the middle layers 
$R^{2}_{22}$ starts increasing from the $I-N_{U}^{'}$ transition itself, but
 a significant increase is observed only at the $N_{U}^{'}-N_{B}$ transition.
The layer-wise $R^{2}_{02}$ and $R^{2}_{20}$ are shown in 
Fig.~\ref{fig:5b}. It is observed that they continue to be relatively insignificant.

The plots of the layer-wise $\theta$  and layer-wise 
$\phi$ are shown in  Fig.~\ref{fig:5c}. Focusing on the data only below
the clearing point, it is observed that as the 
temperature is lowered from the isotropic phase, layer-wise $\theta$ 
values stabilise at certain fixed values in the intermediate phase. The value 
of this angle increases monotonically from $\textit{k}$ = 1 layer to the 
$\textit{k}$ = $d$ layer, clearly indicating that the layer - wise 
primary director (of this phase) bends gradually from the lower substrate to the upper
substrate, corresponding to the expected bent-director structure. The 
layer-wise $\phi$ angles show a sudden flip by approximately
$ 90^{0}$ just below the $ I-N_{U}^{'}$ transition, particularly in the region of 
middle layers, pointing to the onset of the dominant order essentially 
confined to the laboratory YZ plane, as the system
is cooled. The profile of the layer-wise angle  also suggests that this 
bent-director structure is not affected by the $N_{U}^{'} - N_{B}$ transition.

\subsection{Type-B Film}
\begin{figure}
\centering
\includegraphics[width=0.38\textwidth]{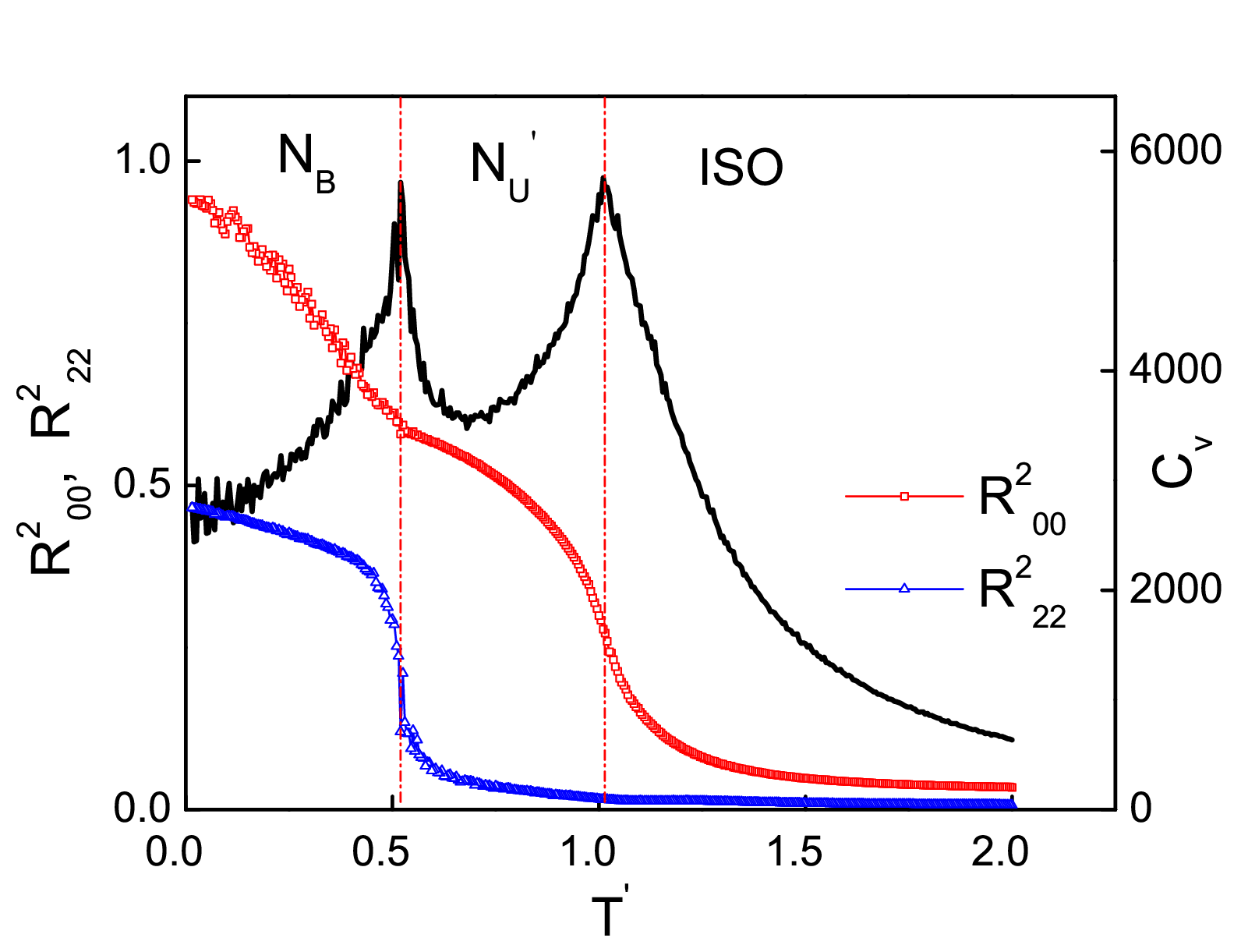}
\caption{Variation with temperature of system properties in Type-B film - peaks
of the specific heat $C_{v}$ mark the phase transitions $I - N_{U}^{'}$ and 
$N_{U}^{'} - N_{B}$. The phases are identified by the growth of the major 
order parameters $R^{2}_{00}$ and $R^{2}_{22}$.}
\label{fig:6}
\end{figure}

The simulations of the Type-B film were carried out similarly
using different anchoring conditions at the substrates, as mentioned earlier.
We note that in this film the interaction between the substrate and the surface
film layers is biaxial in nature, similar to the interaction between  the 
LC molecules in the bulk. This system undergoes two transitions at 
$T_{1}=1.014$, and the second at $T_{2}=0.527$, exhibiting the sequence
$N_{B} - N_{U}^{'} - I$ (see Fig.~\ref{fig:6}).    
Introduction of biaxial coupling between the substrates and the surface film
layers, in the presence of an already constrained bent director of the molecular
 $\textit{z}$-axes, introduces incompatible boundary conditions on the minor 
 axes as well, and seem to lead to interesting manifestations.
 
\begin{figure}
\centering{
\subfigure[]{\includegraphics[width=0.38\textwidth]
{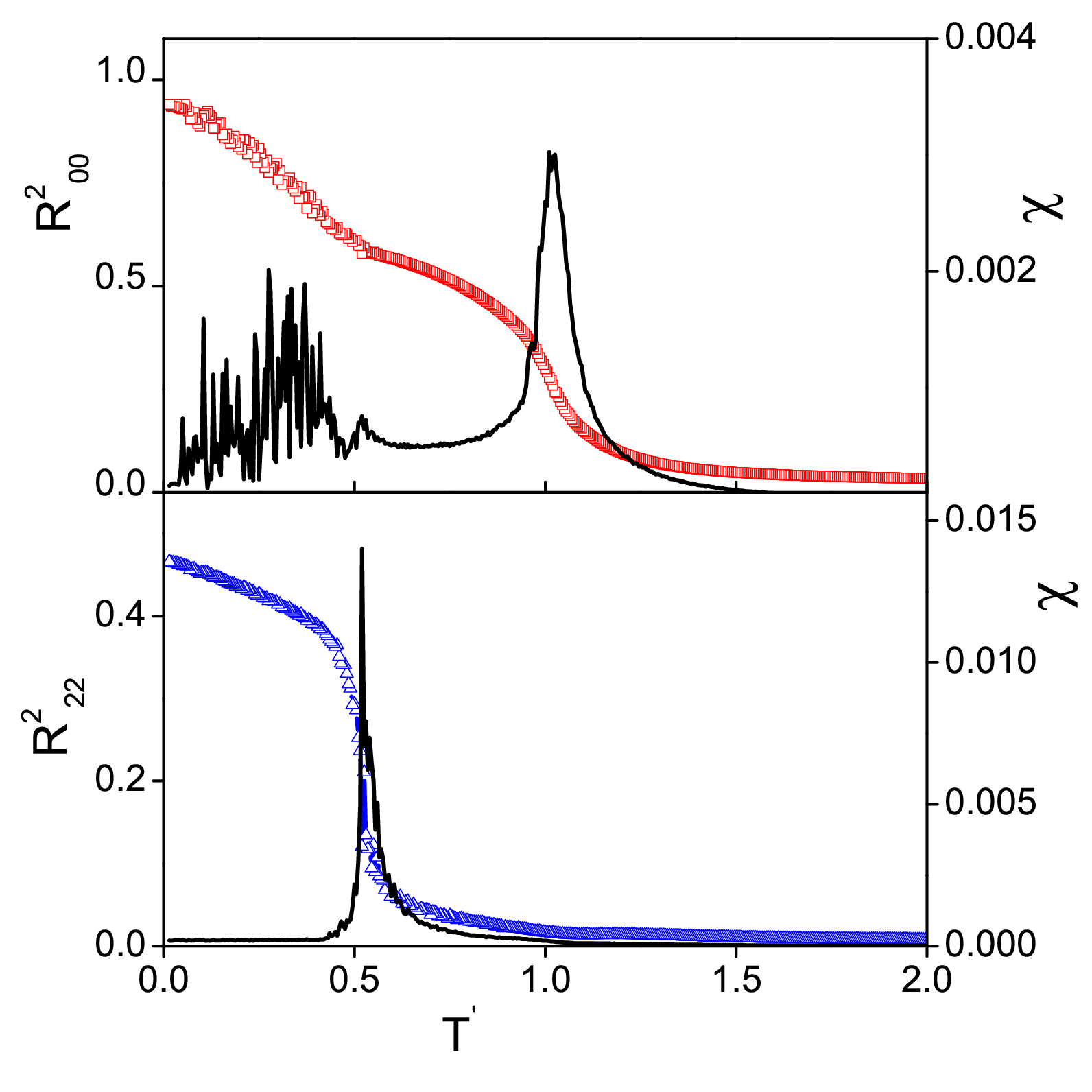}
\label{fig:7a}}
\subfigure[]{\includegraphics[width=0.38\textwidth]
{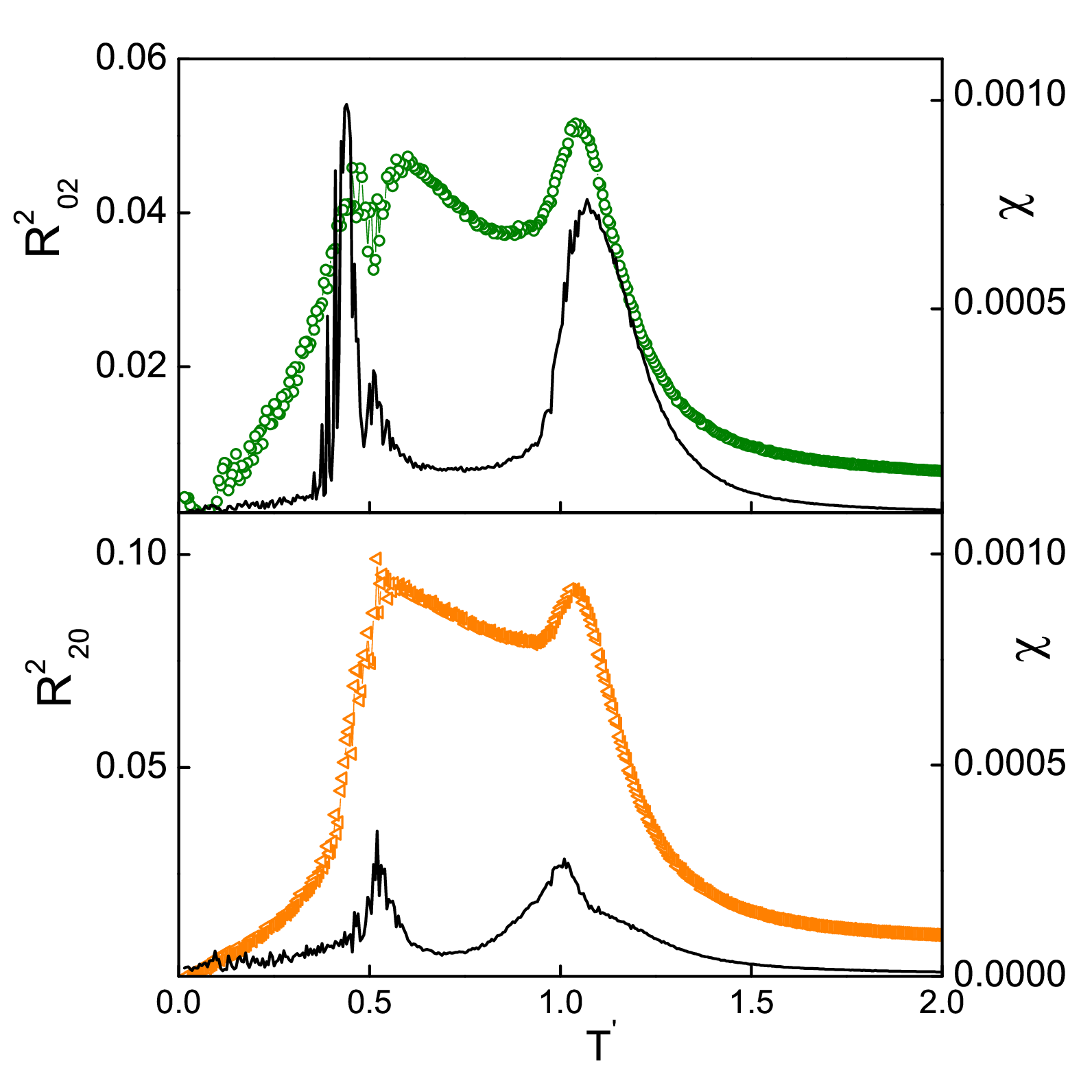}
\label{fig:7b}}
}
\caption{Variation with temperature of the four system order parameters 
 (lines with symbols) and their susceptibilities (continuous lines) in 
Type-B film (a) $R^{2}_{00}$ and $R^{2}_{22}$ (b) $R^{2}_{02}$ and 
$R^{2}_{20}$. }
\label{fig:7}
\end{figure}

Referring to Fig. \ref{fig:6}, it may be noted that the the degree of 
biaxial order in the intermediate phase ($N_{U}^{'}$) is relatively 
smaller than in the same phase of Type-A film (Fig.\ref{fig:3}). 
Figs.\ref{fig:7a} and \ref{fig:7b} show the temperature variation of 
the order parameters and their susceptibilities. Interestingly, 
the dominant order $R^{2}_{00}$ and its susceptibility exhibit 
large fluctuations in the biaxial nematic phase, unlike in Type-A film.

\begin{figure}
\centering{
\subfigure[]{\includegraphics[width=0.38\textwidth]{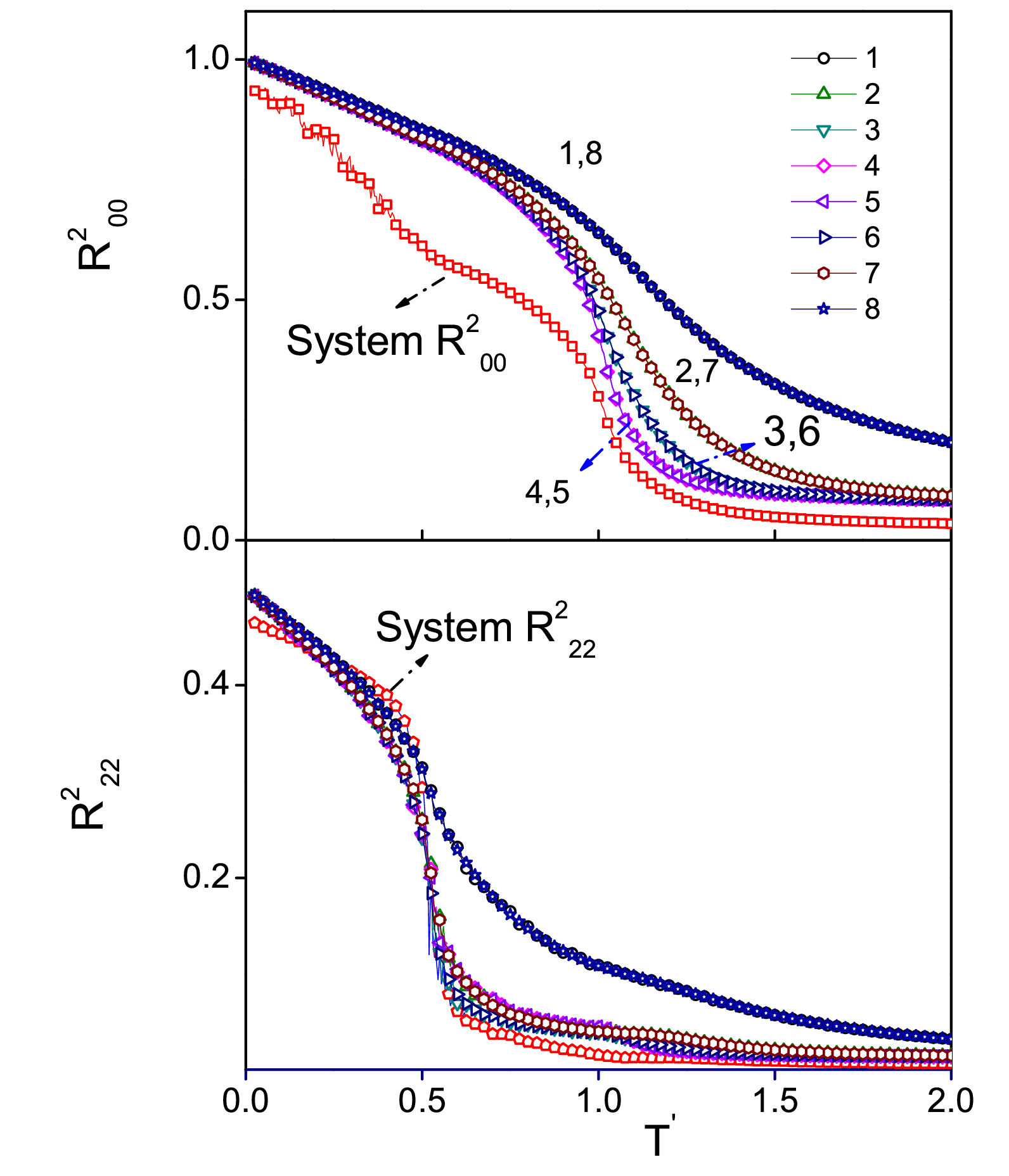}
\label{fig:8a}}
\subfigure[]{\includegraphics[width=0.38\textwidth]{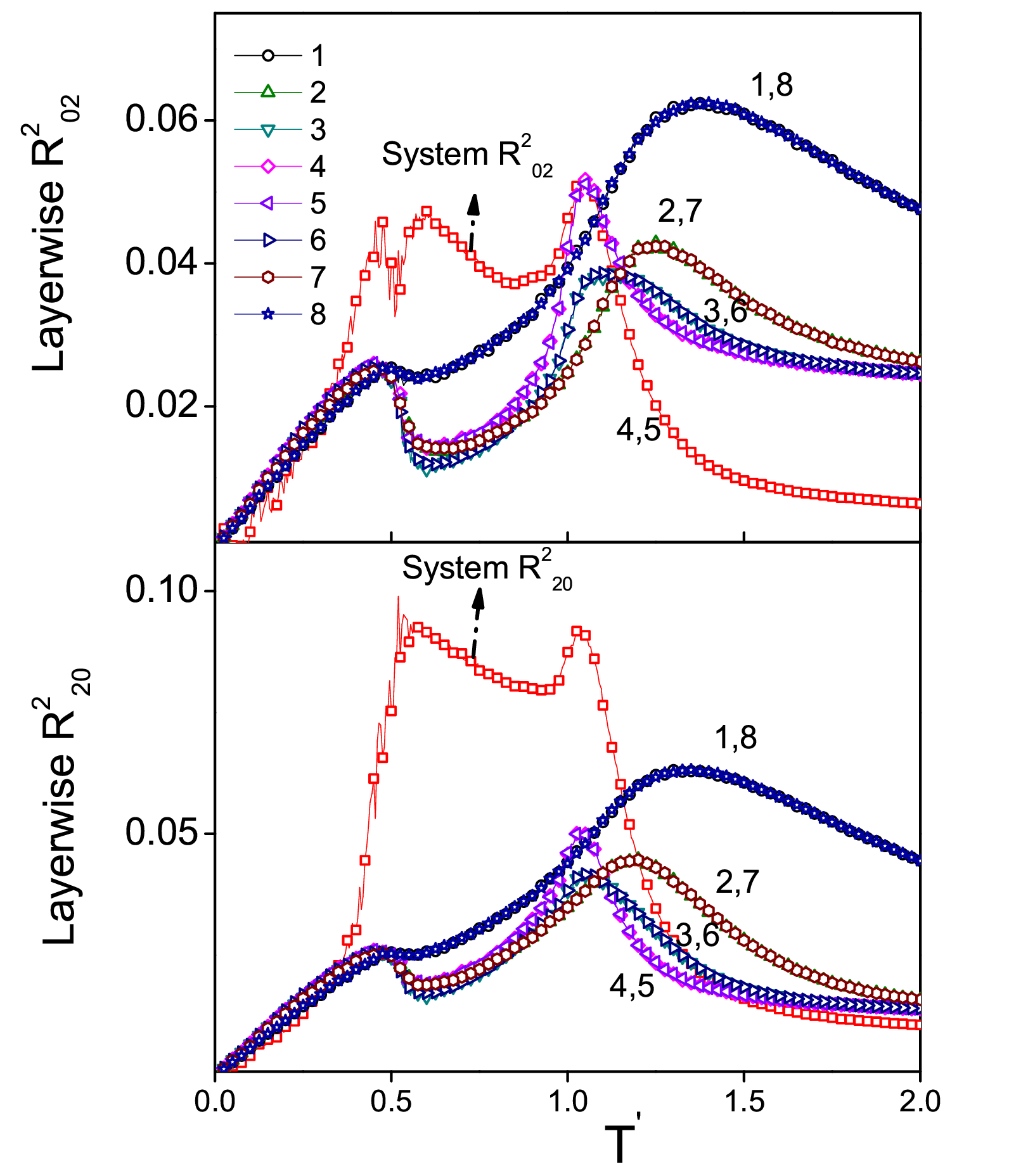}
\label{fig:8b}}
\subfigure[]{\includegraphics[width=0.35\textwidth]{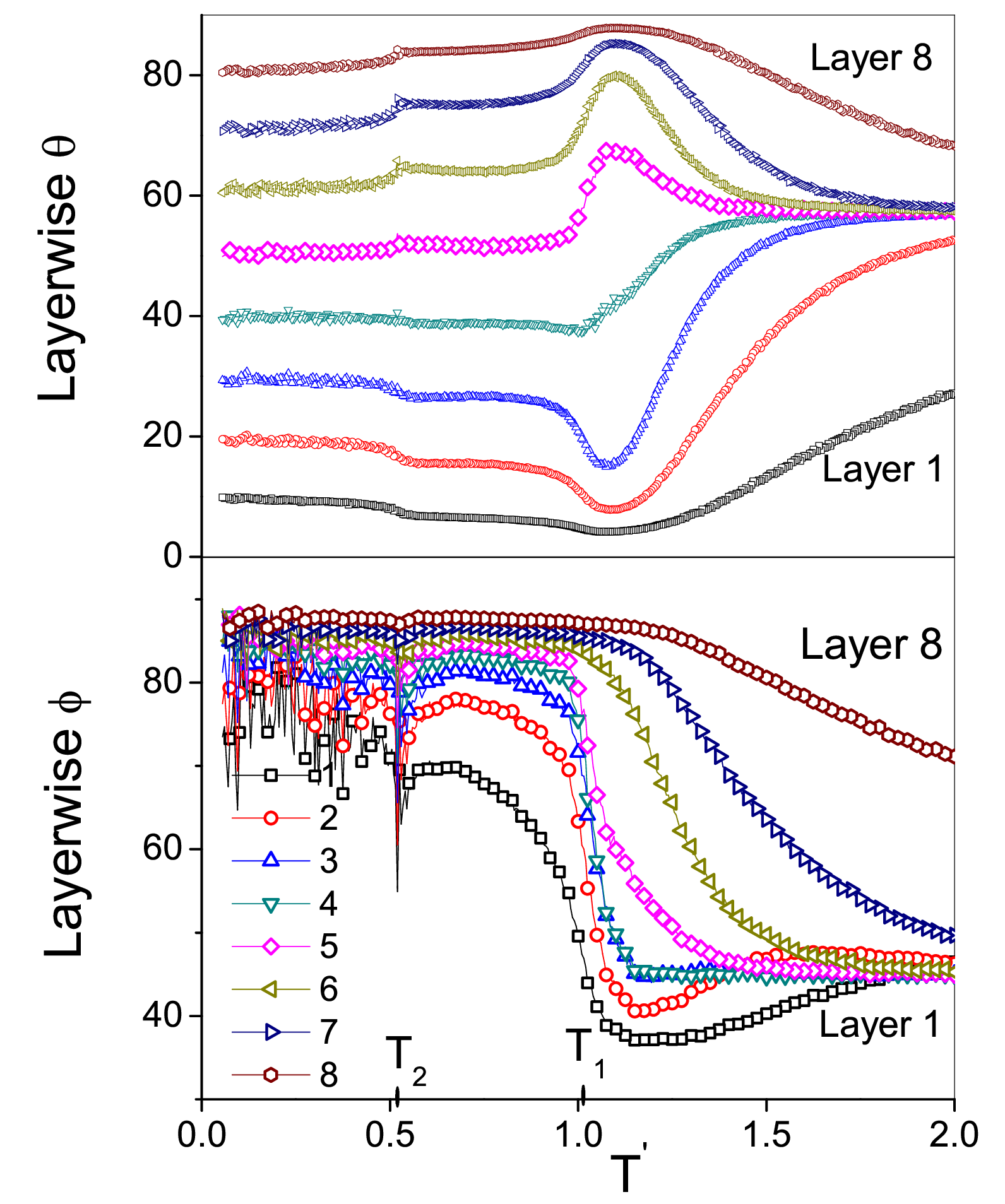}
\label{fig:8c}}
}
\caption{Layer-wise variation of different order parameters of the 
Type-B film, with temperature: (a) $R^{2}_{00}$ and $R^{2}_{22}$ (b) 
$R^{2}_{02}$ and $R^{2}_{20}$ (c) Angles $\theta$ and $\phi$ (see text).}
\label{fig:8}
\end{figure}

The system and layer-wise variations of different order parameters are shown
in Figs.~\ref{fig:8a} - \ref{fig:8b}. The layer-wise $R^{2}_{00}$ values in 
Fig.~\ref{fig:8a} do not fluctuate in the biaxial phase and attain maximum
ordering, interestingly unlike the system order parameter. The layer-wise 
biaxial order shows a non - zero value just below the $I-N_{U}^{'}$ transition 
and grows gradually, while the system biaxial
order develops more appreciably at the $N_{U}^{'}-N_{B}$ transition. Both 
fluctuate significantly in the biaxial phase similar to the  other order 
parameters. The layer-wise $R^{2}_{02}$ and $R^{2}_{20}$ shown in 
Fig.~\ref{fig:8b} have small, nonzero values in the  intermediate phase, they also fluctuate more in the biaxial phase.  

 The layer-wise $\theta$ and $\phi$ values shown in 
 Fig.~\ref{fig:8c} depict variations with temperature in the two nematic 
 phases, which are largely similar to the behaviour of Type-A film, 
 but for significant fluctuations on the onset of the $N_{U}^{'} - N_{B}$
 transition. It again appears that the layer-wise magnitudes of the two dominant
 orders are relatively stable, but their orientations are not. 
 
 Typical (low) errors quoted in Table \ref{tab:table1}, particularly in the case of 
 $R^{2}_{00}$ in biaxial phase of Type-B film, cannot account for its large 
 fluctuations, discernible by its non-smooth variation with temperature
 in this phase. We investigated the possible origin of these fluctuations
 by comparing the results of distinct MC simulations on both the films covering 
 the temperature range, starting with ten different
 initial random configurations. We examined the statistics of the resultant 
 averages (of corresponding quantities in the different phases), arising
 from different trajectories in the configuration space. We find that the JK 
 errors from MC simulations over different trajectories are comparable in 
 each of these films at corresponding temperatures, and as 
 small as indicated in Table \ref{tab:table1}. However the scatter of the MC averages 
 of order parameters, in particular $R^{2}_{00}$, of Type-B film in its 
 biaxial nematic phase, is much larger than any of its single trajectory 
 JK estimates.  The standard deviation of the 
 average $R^{2}_{00}$ value in this phase of Type-B film, obtained from 
 the different trajectories is typically $1 \times 10^{-3}$, large compared to the
 single trajectory  JK error of $1 \times 10^{-4}$. This points to the 
 scenario that the biaxial nematic phase of
 Type-B film does not host a single unique free energy minimum, but is rather
 shallow with many local minima, to which each of the trajectories is 
 attracted depending on the initial conditions. This seems to account also 
 for the fluctuation of the order parameters (and their susceptibilities) 
 in this phase. We will return to this point later in the text.

 \begin{figure}
 \centering{
\subfigure[]{\includegraphics[width=0.38\textwidth]{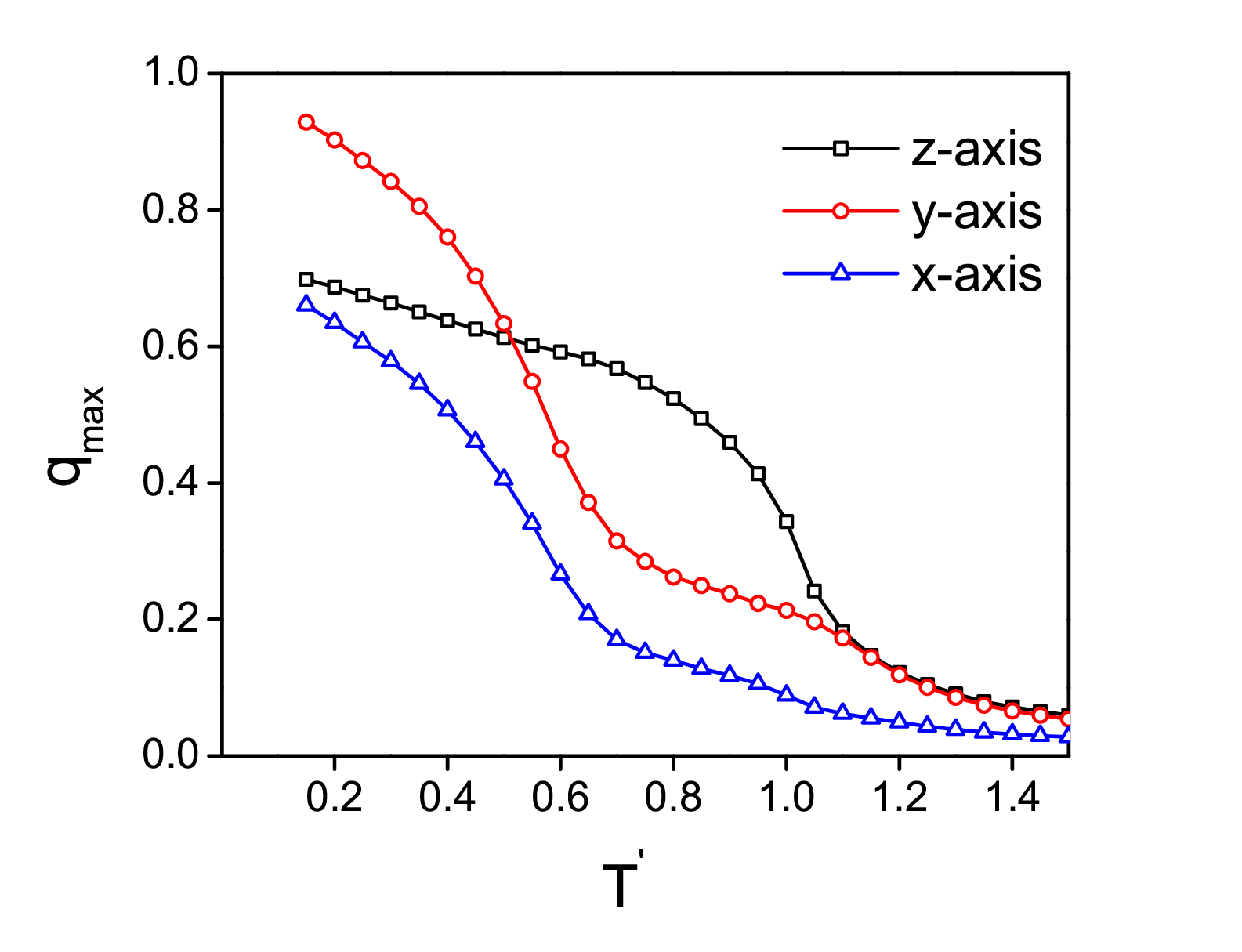}
\label{fig:9a}}
\subfigure[]{\includegraphics[width=0.38\textwidth]{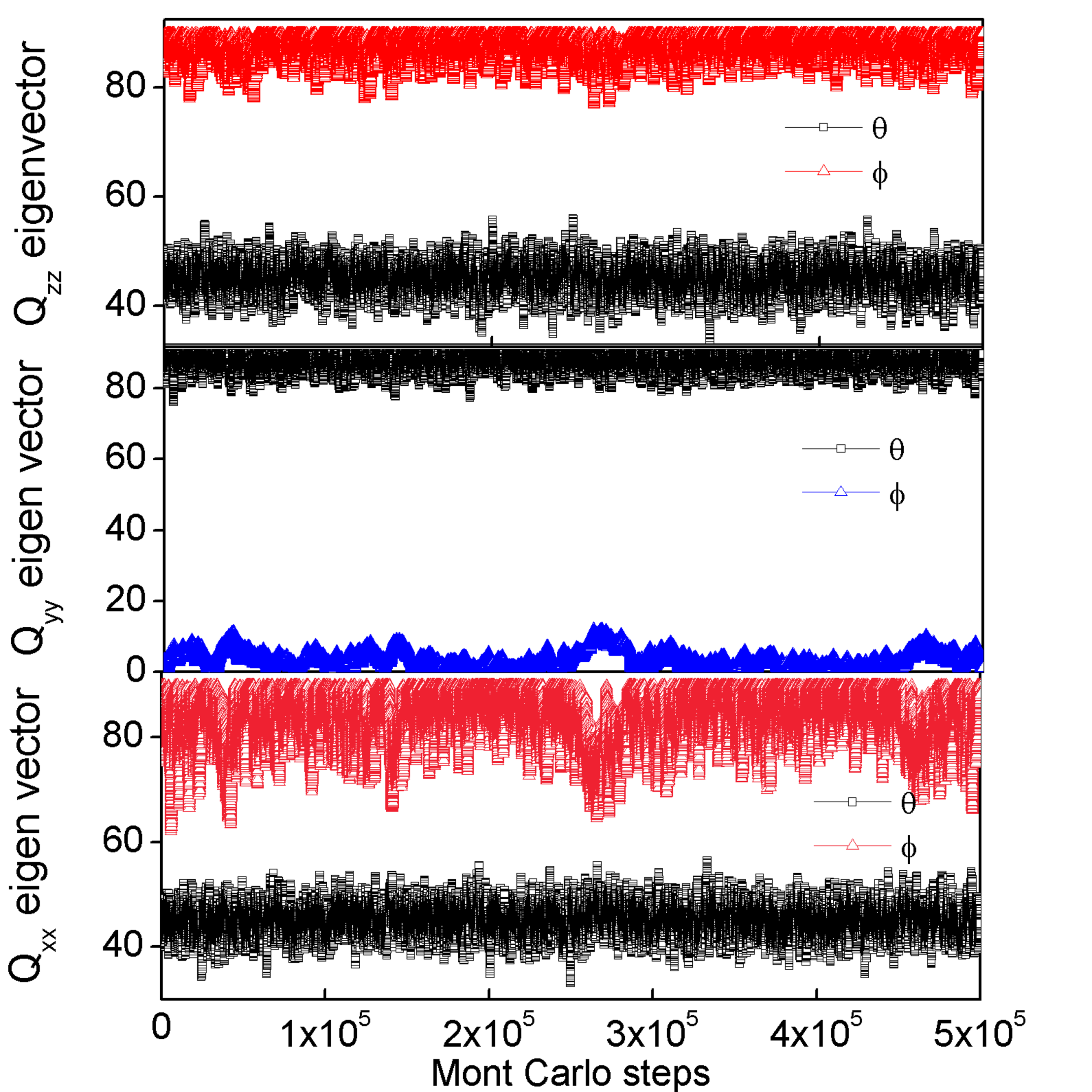}
\label{fig:9b}}
}
\caption{ Type-A film : (a) Variation of maximum eigen values ($q_{max}$) 
associated with each of the ordering tensors of the three molecular axes 
($\textit{x},  \textit{y}, \textit{z}$), as a function of temperature; 
(b) The orientations  of the corresponding eigen vectors plotted as a 
function of the Monte Carlo steps after equilibration, at temperature 
$T=0.5$ below the $N_{U}^{'} - N_{B}$  transition temperature. }
 \label{fig:9}
\end{figure}

 \begin{figure}
 \centering{
\subfigure[]{\includegraphics[width=0.38\textwidth]{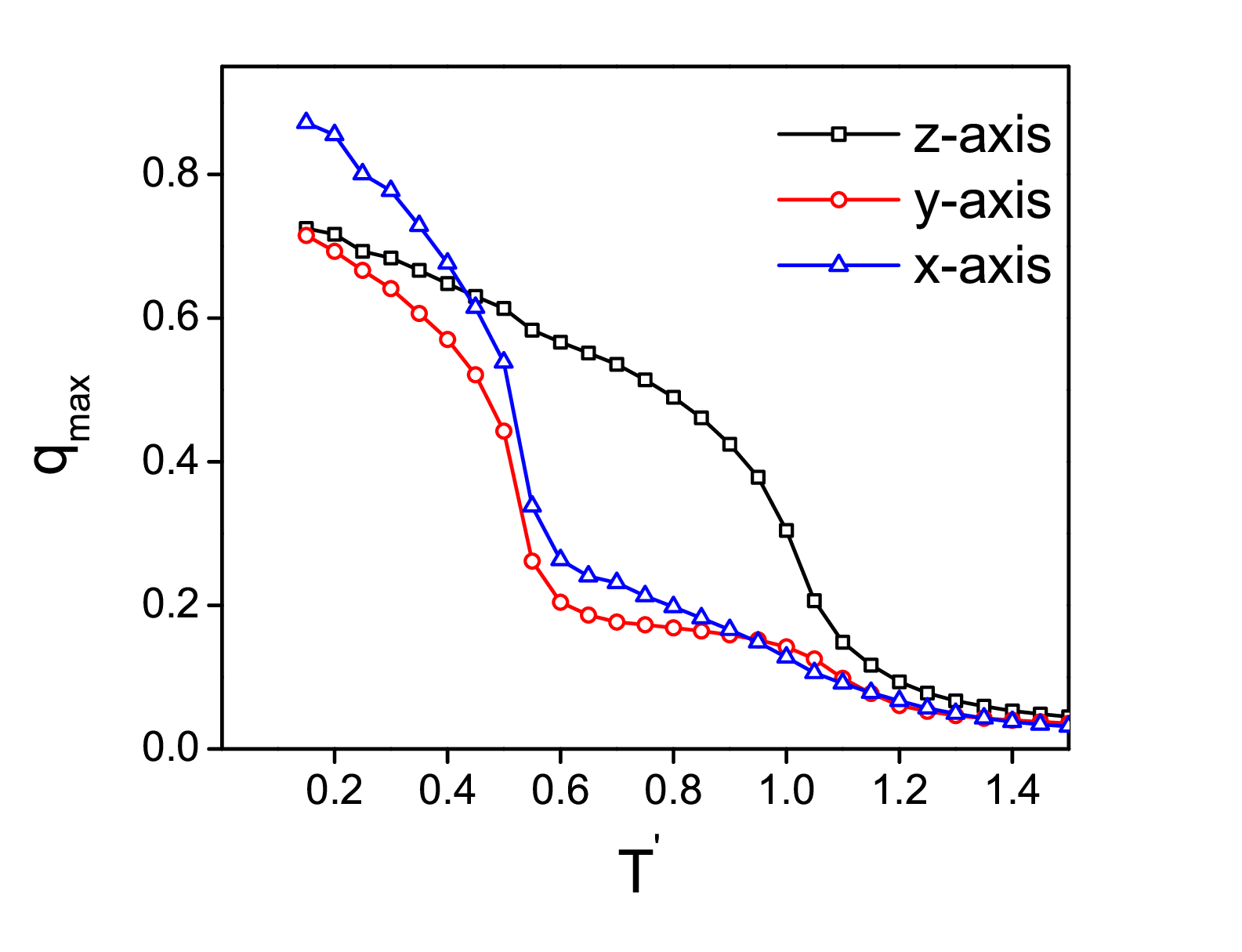}
\label{fig:10a}}
\subfigure[]{\includegraphics[width=0.38\textwidth]{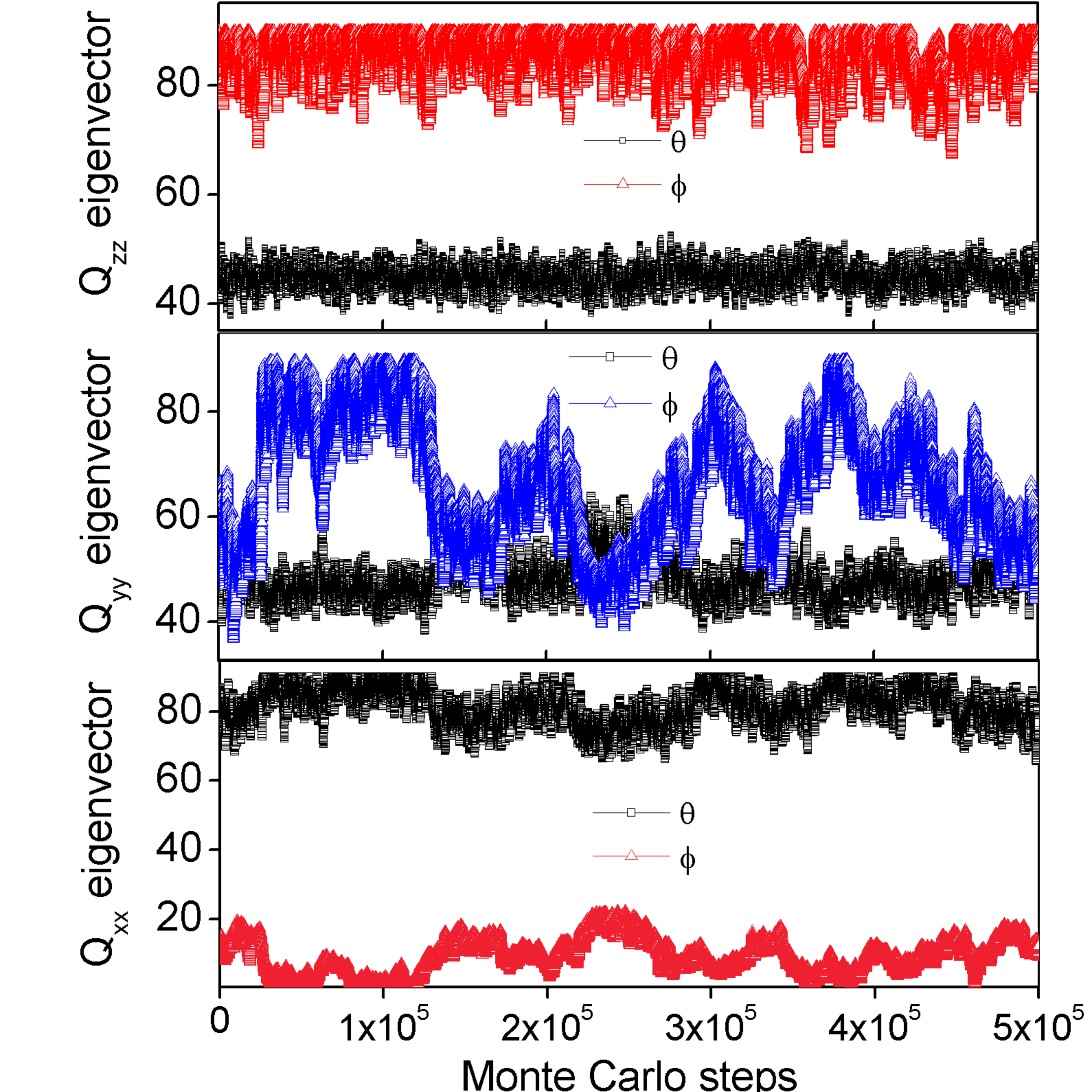}
\label{fig:10b}}
}
\caption{ Type-B film : (a)Variation of maximum eigen values ($q_{max}$) 
associated with each of the ordering tensors of the three molecular axes 
($\textit{x},  \textit{y}, \textit{z}$), as a function of temperature; 
(b) The orientations  of the corresponding eigen vectors plotted as a 
function of the Monte Carlo steps after equilibration, at temperature 
$T=0.5$ below the $N_{U}^{'} - N_{B}$  transition temperature. }
 \label{fig:10}
\end{figure}

We now discuss the evolution of the sample uniaxial order $(R^{2}_{00})$ in 
the two films as the system transits through the two transitions, specifically in 
comparison with the layerwise behaviour (Figs.~\ref{fig:5a}, \ref{fig:8a}).
Focussing on Type-A film initially, we note that the sharp increase in this order 
at the $N_{U}^{'} - N_{B}$ transition is unlike the variation in the 
bulk sample (Fig.~\ref{fig:2}), and is not supported by its layer-wise variations. 
The latter are smoothly varying across the transition, and the director angles
$(\theta, \phi)$ also do not betray the presence of an intermediate phase 
transition. And yet, the bulk order of the film, defined and computed
as corresponding to the maximum eigen value of the ordering tensors of
the three molecular axes, is very sensitive to this transition. The case
of the Type-B film is qualitatively the same, but for the onset of 
significant fluctuations in the $N_{B}$ phase at $T_{2}$. This points 
to the need to further investigate this system in terms of thermal
variations of the eigen values of the three axes separately and examine
their behavior across this transition. 

           Accordingly, we computed equilibrated averages (over the film)
of the maximum eigen values ($q_{x}$, $q_{y}$, $q_{z}$) of the ordering 
tensors ($Q_{xx}, Q_{yy},Q_{zz}$) of the three molecular axes and we depict 
their variation for the case of Type-A film in Fig.~\ref{fig:9a}. We also 
show the directions ($\theta, \phi$) of the 
corresponding three eigen vectors, as a function of MC steps after 
equilibration, in the $N_{B}$ phase, in Fig.~\ref{fig:9b}. From these 
two figures, it is 
evident that the onset of the biaxial phase in this confined system
(at $T_{2}$=0.5) leads to a sudden change in the direction
of dominant order of the film itself. While the alignment of the long 
molecular axes defines the primary director till $T_{2}$, it is the 
molecular $\textit{y}$ axes which are the most ordered among the three,
below this temperature. In conjunction with Fig.~\ref{fig:9b}, we see that 
the ordering direction of this axis is indeed in the laboratory 
X-direction in the biaxial phase, while the other two eigen vectors are 
confined to the laboratory YZ plane, mutually perpendicular to each other. 
It may be noted that the onset of a biaxial phase thus leads to maximal 
ordering of the second major axes, wholly contained in the plane of the 
substrate, and the anchoring conditions imposed in this case constrain
the molecular $\textit{x}$-axes and $\textit{z}$-axes due to anchoring effects, leaving the 
$\textit{y}$-axis to freely develop significant order in the plane of the film. 

             In contrast, Type-B film which imposes anchoring restrictions 
on the minor axes of molecules as well at both the ends, presents 
a very different scenario. We refer to Fig.~\ref{fig:10a} showing ($q_{x}$,
$q_{y}$, $q_{z}$) as a function of temperature in this film. At the
onset of the second transition at $T_{2} \sim 0.51$, $q_{z}$ ($\sim$ 0.617) 
is higher than the $q_{x}$ ($\sim$ 0.536) and $q_{y}$ ($\sim 0.447$).  
However on further cooling, $q_{x}$ crosses 
the value of $q_{z}$ ($\sim 0.632$ at $T^{'}$ = 0.433), while $q_{y}$ 
remains less than $q_{z}$. At lower temperatures $q_{x}$ saturates at $\sim$ 0.87, 
while $q_{y}$ and $q_{z}$ saturate to a value of $\leq$ 0.72. Eigen vector of 
$q_{z}$ makes an angle $\theta \sim 45^{0}$ with the laboratory Z-direction and 
$\phi \sim 90^{0}$ with laboratory X-direction, thereby indicating that 
the bent-director structure originating from the ordering of the molecular 
$\textit{z}$-axes is contained in the YZ plane, as is also the case in the 
high temperature nematic phase. Curiously the eigen vector of $q_{y}$
is oriented at angles $\theta \sim 45^{0}$ and with a fluctuating $\phi$ 
varying between $0^{0}$ to $90^{0}$ (see Fig.~\ref{fig:10b}). The maximal 
ordering direction in this film is determined by the ordering tensor of 
the molecular $\textit{x}$-axes and its azimuthal angle $\phi$ is 
eventually contained in the plane of the substrate pointing to the 
laboratory X-direction. It may be noted that the corresponding fluctuations 
of the azimuthal angle $\phi$ of the local directors of the $\textit{x}$ 
and $\textit{y}$ molecular axes are complementary (Fig.~\ref{fig:10b}). 

               It is now clear that the observed significant fluctuations
in the thermal averages, particularly of $R^{2}_{00}$
at the onset of the $N_{B}$ phase, arises due to the fluctuations in the 
 directions of the different ordering tensors (Figs.~\ref{fig:7a} and \ref{fig:7b}).
 The qualitatively different scenario of the Type-B film, 
relative to Type-A, seems to arise due to the imposition of additional 
restrictions on all the molecular axes at the two substrates. 

       Comparing the two films in their biaxial phase, we observe that
both have primary director (defined as the direction of maximum molecular 
order) contained in the plane of the substrates, with molecular 
$\textit{y}$-axes defining such a direction for the  Type-A film while 
$\textit{x}$-axes play that role for Type-B film. However, imposition of 
anchoring constraints on all molecular axes leaves the second film frustrated, 
leading to significant fluctuations of the ordering directions. It is in this
context that the earlier observation on the scattering of the MC average
values arising from different initial random configurations points to 
a glass-like behaviour of the frustrated Type-B film, once the onset of its
biaxial phase takes place.

From a practical point of view, it is simpler to prepare substrates 
which need to restrain only one type of  axes of the
system. Thus Type-A lends itself as a possible stable structure containing
 significant order within the plane of the film, with potential applications.

\subsection{Effect of thickness } 
     
\begin{figure}
\centering
\includegraphics[width=0.38\textwidth]{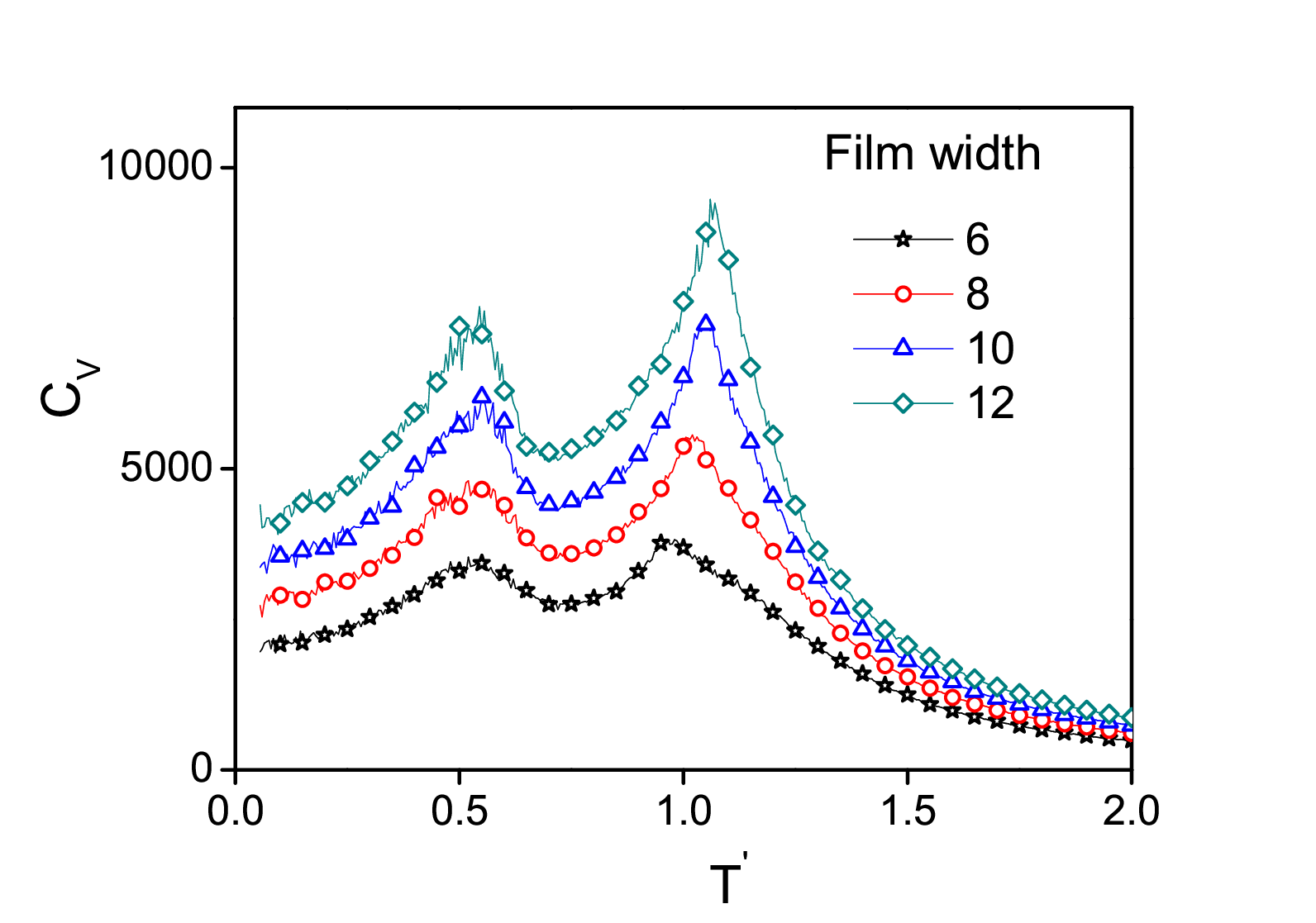}
\caption{ Comparison  of variations of the specific 
heat $C_{v}$ as a function of temperature in films of thickness 
d (in lattice units) = 6 (stars), 
8 (circles), 10 (upward triangles) and  12 (diamonds), in Type-A film. }
\label{fig:11}
\end{figure}

\begin{figure}
\centering
\includegraphics[width=0.5\textwidth]{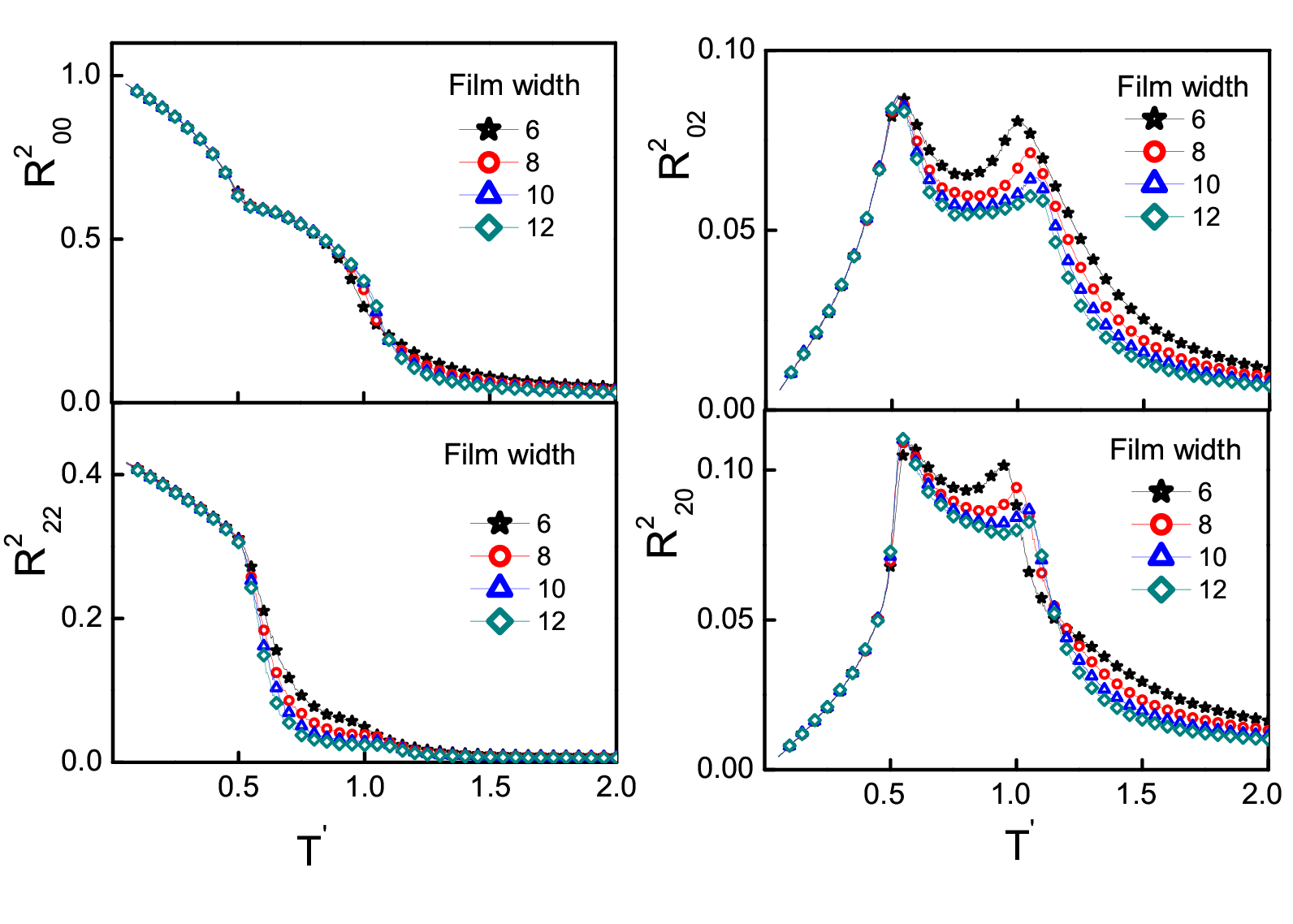}
\caption{ Comparison  of variation of order parameters with
temperature for different film thickness ($\textit{d}$)
in Type-A film; 
$\textit{d}$ = 6 (stars), $\textit{d}$ = 8 (circles), $\textit{d}$ = 10 
(upward triangles) and $\textit{d}$ = 12 (diamonds) } 
 \label{fig:12}
\end{figure}

\begin{figure}
\centering
\includegraphics[width=0.38\textwidth]{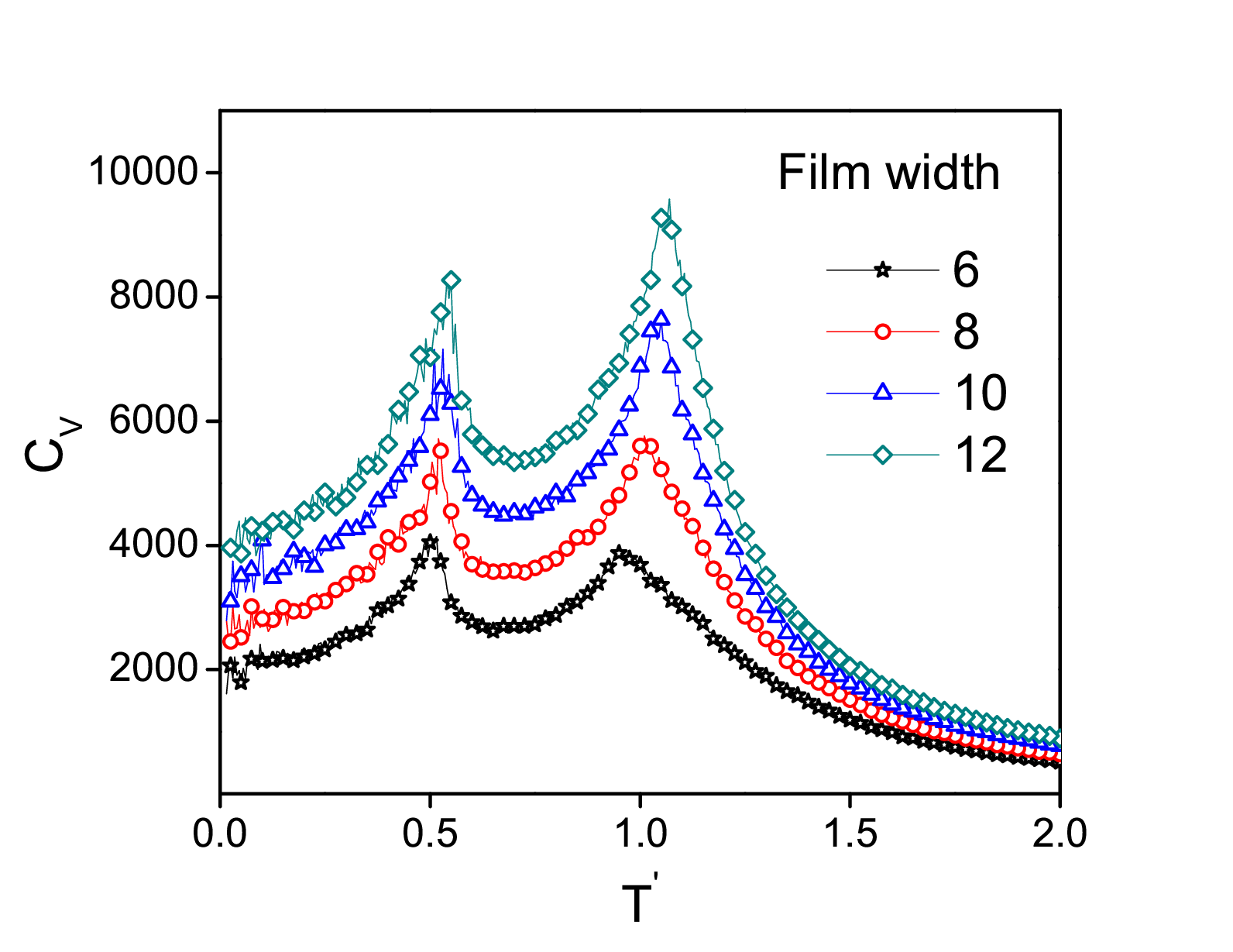}
\caption{ Comparison  of variations of the specific 
heat $C_{v}$ as a function of temperature in films of thickness 
d (in lattice units) = 6 (stars), 
8 (circles), 10 (upward triangles) and  12 (diamonds), in Type-B film. }
\label{fig:13}
\end{figure}

\begin{figure}
\centering
\includegraphics[width=0.5\textwidth]{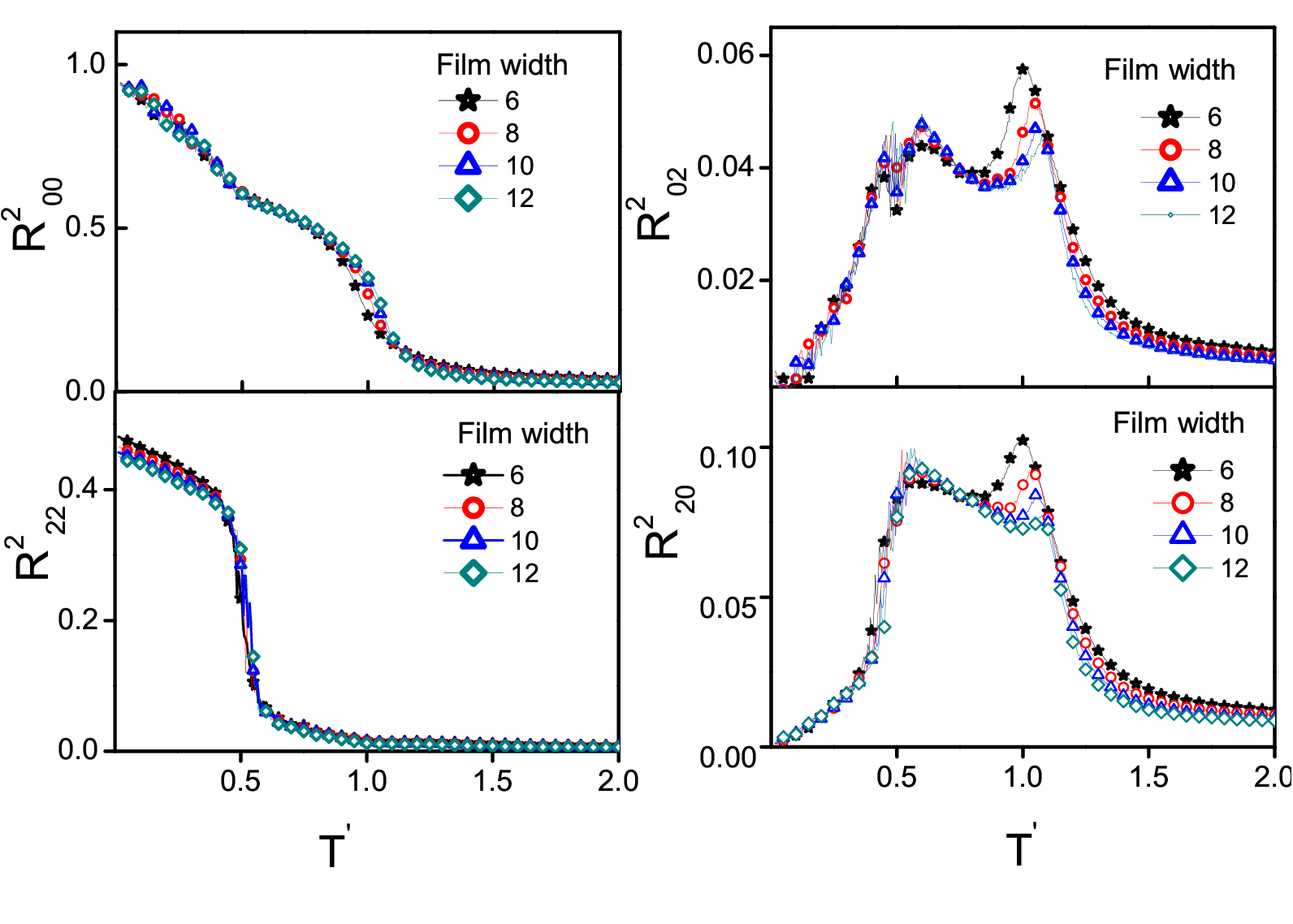}
\caption{ Comparison  of variation of order parameters with
temperature for different film thickness ($\textit{d}$)
in Type-B film.; $\textit{d}$ = 6 (stars), $\textit{d}$ = 8 (circles), 
$\textit{d}$ = 10 (upward triangles) and $\textit{d}$ = 12 (diamonds) } 
 \label{fig:14}
\end{figure}

We examined the 
effect of varying the thickness $'\textit{d}'$ in both films, 
($\textit{d}$ =  6, 8, 10, 12 lattice units), while retaining the same 
lateral dimensions, and strong anchoring conditions at the two substrates.
Figs.~\ref{fig:11} and  \ref{fig:12} show this effect on the specific heat $C_{v}$ 
and the order parameters $R^{2}_{00}$, $R^{2}_{22}$, $R^{2}_{02}$ and $R^{2}_{20}$
in the Type-A film, while  Figs.~\ref{fig:13} and \ref{fig:14} depict the 
variations for a Type-B film, respectively.

\subparagraph*{\textit{Type-A film}}
It is observed from Fig.\ref{fig:11} that the specific heat profiles 
of the Type-A film become sharper as the thickness increases (size effect). 
The order parameter variations shown in Fig.\ref{fig:12} show that the 
uniaxial order parameter $R^{2}_{00}$ increases marginally in the uniaxial 
phase as the  thickness increases, retaining its bent director structure for 
 all thicknesses. The biaxial order $R^{2}_{22}$ shows
 a slight decrease as the thickness increases and shows a marked 
 increase in the temperature of the nematic phase itself for a thinner 
film ($d$=6). The $R^{2}_{02}$ and $R^{2}_{20}$ values 
 decrease as the thickness increases. These observations
indicate that the gross features of director structures in Type-A film are 
relatively insensitive to the film thickness, alluding to some degree of 
flexibility in its design.
         
 \subparagraph*{\textit{Type-B film}} 
 
      The effect of varying the thickness of the Type-B film on the
specific heat profiles (Fig.~\ref{fig:13}) and the order parameters 
(Fig.~\ref{fig:14}) is similar to Type-A film, except the biaxial order. 
The variation of $R^{2} _{22}$ is independent of the film thickness 
unlike the other order parameters. The data in the biaxial phase of 
this film however suffer from large fluctuations at all thickness values.

 \subsection{Effect of anchoring strength}
 
\begin{figure}
\centering
\includegraphics[width=0.5\textwidth]{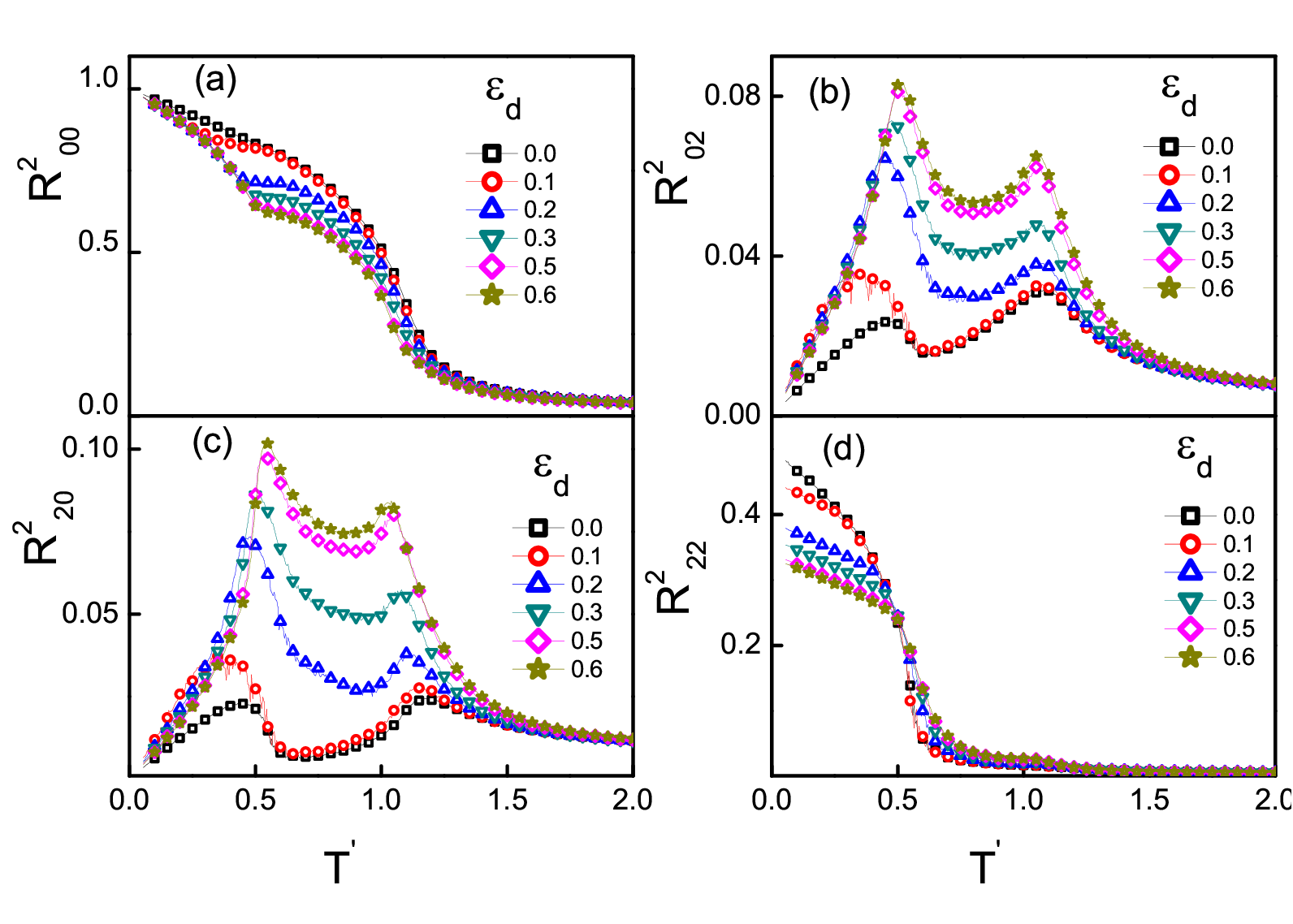}
\caption{ Variation of order parameters  with temperature for different 
anchoring strengths ($\epsilon_{d}$) in Type-A film of thickness 8 lattice units: 
(a) $R^{2}_{00}$ (b) $R^{2}_{02}$ (c) $R^{2}_{20}$ and (d) $R^{2}_{22}$.}
\label{fig:15}
\end{figure}

\begin{figure}
\centering
\includegraphics[width=0.5\textwidth]{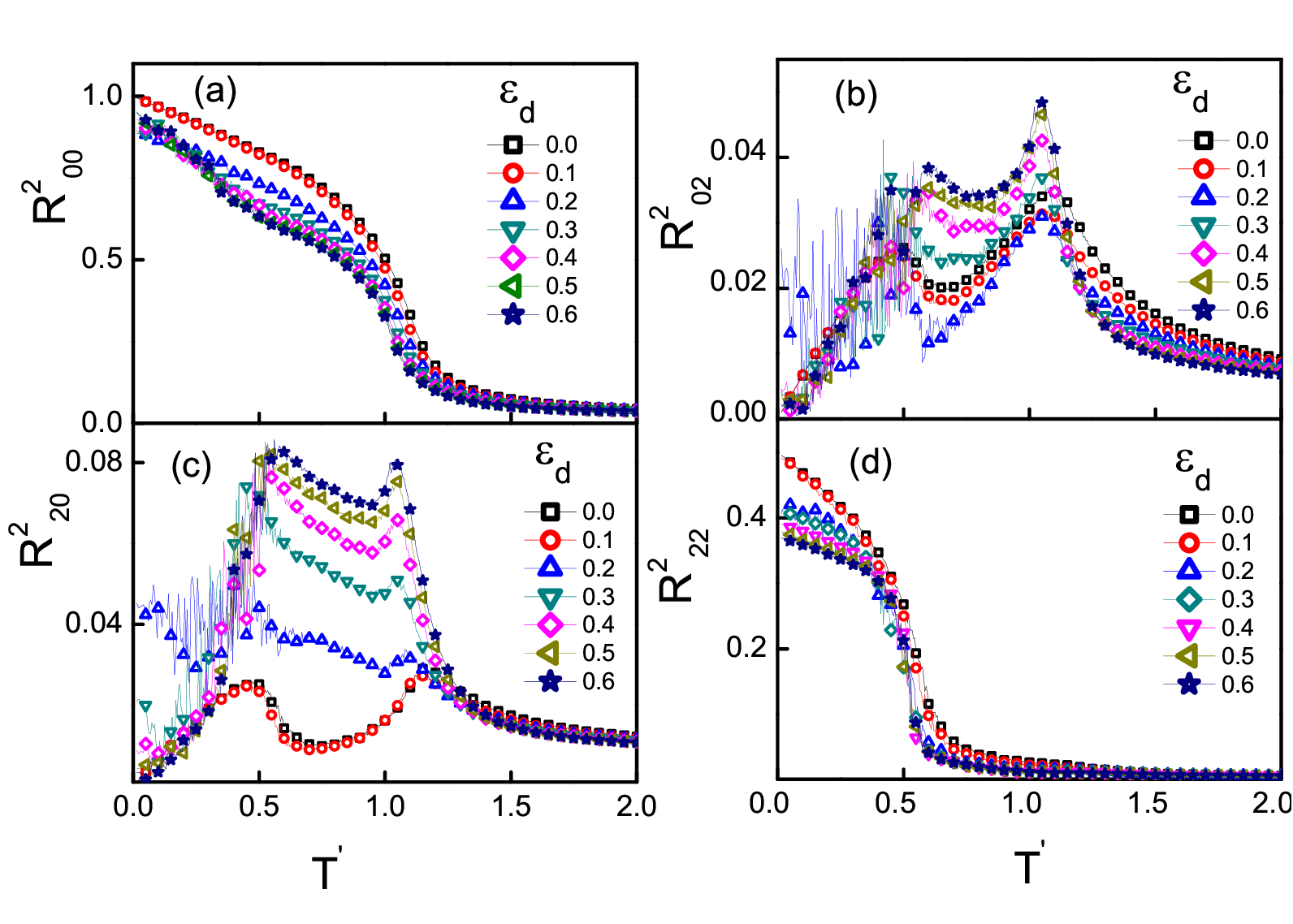}
\caption{ Variation of order parameters  with temperature for different 
anchoring strengths ($\epsilon_{d}$) in Type-B film of thickness 8 lattice units: 
(a) $R^{2}_{00}$ (b) $R^{2}_{02}$ (c) $R^{2}_{20}$ and (d) $R^{2}_{22}$.}
\label{fig:16}
\end{figure}

We further examined both the films (at fixed thickness $d$ = 8) with
respect to a change in the anchoring strength $\epsilon_{d}$ at the 
surface layer. We relax the strong anchoring condition ($\epsilon_{d}$ = 1)
and vary its value now from 0.0 to 0.6 in steps of 0.1, and compute
equilibrium averages (as a function of temperature) of the four order
parameters. We depict their dependence on $\epsilon_{d}$ in
Figs.~\ref{fig:15} and \ref{fig:16} for Type-A and Type-B films,
respectively. The other simulational conditions remain the same as before.
  
We observe from Fig.~\ref{fig:15}(a) and \ref{fig:15}(d) that for anchoring
strengths $\epsilon_{d}$=0.0 (no anchoring influence at the top substrate) 
and for a low value of $\epsilon_{d}$ = 0.1, the uniaxial order $R^{2}_{00}$
and the biaxial order $R^{2}_{22}$  attain maximum values of 1.0 and 0.5, 
respectively, - for example without the characteristic features observed 
earlier (with $\epsilon_{d}$ = 1) at the onset of the $N_{B}$ phase.
 Their variations are similar to bulk LC systems without confining surfaces
 (see Fig.~\ref{fig:2}). 
For $\epsilon_{d}\geq 0.2$ the primary director assumes a bent structure 
and the primary order is constrained with an upper bound of 0.8 for 
$\epsilon_{d}=0.2$ in the uniaxal nematic phase. A similar sharp difference is 
exhibited by $R^{2}_{22}$ as well, and its temperature 
variation qualitatively changes above the threshold value of 
$\epsilon_{d}=0.2$. The order parameters  $R^{2}_{02}$ and $R^{2}_{20}$, 
shown in Fig.~\ref{fig:15}(b) and (c), start with very low values at $\epsilon_{d}$
 = 0.0 and 0.1, but increase significantly for higher anchoring strength 
($\epsilon_{d}\geq 0.2$). Thus it appears that a minimum threshold 
anchoring strength ($\epsilon_{d} \geq 0.2$) is necessary for the film 
to exhibit the curious structures, reported earlier, arising from the 
bent formation of its director. The progressive development of this 
scenario with increase of $\epsilon_{d}$ is evident from the gradual
decrease of the system primary order  (Fig.~\ref{fig:15}(a))
in the uniaxial phase.

         Fig.~\ref{fig:16} depicts the anchoring transition in a Type-B film.
 It is to be noted (Fig.~\ref{fig:16}(a)) that though the anchoring transition
 is observed for $\epsilon_{d}\geq 0.2$, the primary director displays a 
 prominent bent director structure for  values of $\epsilon_{d} \sim 0.6$. The 
 variation of $R^{2}_{22}$, $R^{2}_{02}$ and $R^{2}_{20}$ (shown in Figs.~\ref{fig:16}
 (b) - Fig.~\ref{fig:16}(d)) is similar to that 
 for the Type-A film, except that fluctuations are large at the anchoring 
 transition, especially for $R^{2}_{02}$ and $R^{2}_{20}$.
 
\section{Conclusions}
Equilibrium director structures in thin planar films of biaxial LC medium, 
imposing strong hybrid anchoring conditions at the two substrates on the molecular 
long axes ($z$-axes), are investigated through Monte Carlo simulation,  
(Boltzmann sampling) based on a Hamiltonian model under dispersion
approximation (at $\lambda_{d}$ = 0.35). Geometrical confinement induces a
small degree of biaxial order in the intermediate phase, whereas unconfined bulk
system in this phase has uniaxial symmetry. We refer to this intermediate phase,
which is essentially uniaxial with inhomogeneous distribution of biaxial order
as $N_{U}^{'}$.  Type-A film is coupled to the 
substrate only via the molecular long axes ($\textit{z}$-axes), while in Type-B film
the LC molecules interact with the substrate involving the three molecular axes.
Detailed MC simulational studies are carried out on these films covering
both the mesophases, focussing on the temperature variation of the 
director structures, both of bulk film as well as its layers. A comparative analysis
of the data shows that the primary director, defined by the ordering of 
the molecular $z$-axes and constrained in a plane perpendicular to the 
substrates, changes its role as the direction of maximal order qualitatively. 
In the biaxial phase, the maximum orientational
ordering arises from the minor molecular axes ($\textit{y}$-axes in the case of Type-A
film and $\textit{x}$-axes in the case of Type-B film). This order is contained
in the plane of the substrates, and it develops towards its unhindered 
maximum value of unity as the system is cooled in the biaxial phase. 
This shows that geometrical constraints, and hence consequent averaging 
over the layers, are no more at play in this phase.
We also find large fluctuations of this major order in Type-B film, 
unaccounted by typical MC error estimates (exceeding by an order of magnitude).
 Our further studies with multiple initial random configurations,
show that Type-B film  perhaps hosts shallow free energy minimum with 
several local minima (like glass structure), and hence forced to  
giving rise to non-unique MC averages depending
on the trajectory in the configuration space. This is clearly  due to the 
frustration induced in the Type-B system under the simultaneous influence of 
hybrid anchoring conditions affecting the minor axes on one hand, and the 
ordering effect on them due to the Hamiltonian at the onset of the 
biaxial phase on the other. These are 
borne out by the observed fluctuations in this film in the angles of the primary
director of different layers. Such effects are not seen in Type-A film, which
displays a unique MC average value, independent of the trajectory of the 
MC evolution. We conclude that the Type-A film has a very stable 
in-plane dominant ordering, in comparison to Type-B film. Further, its
anchoring conditions, imposing only on the molecular long axes, are more readily
realizable in a laboratory. Both the systems are rather insensitive to the 
film thicknesses. We also find that the relative variation of anchoring strengths
plays a role in inducing the above confinement effects. The anchoring induced 
transition observed while tuning these parameters
suggests that a minimum threshold anchoring strength is necessary to realize these
in-plane structures in the biaxial nematic phases.

\section{Acknowledgments}
 The simulations are carried out in the Centre for Modelling Simulation 
 and Design (CMSD) at the University of Hyderabad. BKL acknowledges the 
 financial support of University Grants Commission 
 of India for the grant of a research fellowship.

\end{document}